\documentclass[preprint,12pt]{elsarticle}
\usepackage{amssymb}

\usepackage{color}

\journal{JQSRT}

\newcommand{\hh}{$\,{\rm H}_2\,$}
\newcommand{\dd}{$\,{\rm D}_2\,$}
\newcommand{\al}{\alpha}
\newcommand{\be}{\beta}
\newcommand{\pa}{\partial}

\newcommand{\Si}{\Sigma}

\newcommand{\tha}{\theta}

\newcommand{\rar}{\rightarrow}

\begin{document}

\begin{frontmatter}

\title{The hydrogen molecule $\rm{H}_{2}$ in inclined configuration in a weak magnetic field}

\author{Alexander Alijah}

\address{Universit\'e de Reims Champagne-Ardenne, Groupe de Spectrom\'etrie Mol\'eculaire et Atmosph\'erique
(UMR CNRS 7331), U.F.R. Sciences Exactes et Naturelles,\\ Moulin de la Housse B.P. 1039, F-51687
Reims Cedex 2, France}
\ead{alexander.alijah@univ.reims.fr}

\author{Juan Carlos~L\'opez Vieyra}
\ead{vieyra@nucleares.unam.mx}

\author{Daniel~J.~ Nader}
\ead{daniel.nader@correo.nucleares.unam.mx}

\author{Alexander V.~Turbiner}
\ead{turbiner@nucleares.unam.mx}

\address{Instituto de Ciencias Nucleares, Universidad Nacional Aut\'onoma de M\'exico,\\ Apartado Postal 70-543,
04510 Ciudad de M\'exico, M\'exico}

\author{H\'ector~Medel~Cobaxin}
\ead{hjmedel@gmail.com}
\address{Instituto Tecnol\'ogico de Estudios Superiores de Monterrey,\\
64849 Monterrey, N.L., M\'exico}

\begin{abstract}
Highly accurate variational calculations, based on a few-parameter, physically adequate
trial function, are carried out for the hydrogen molecule \hh in inclined configuration,
where the molecular axis forms an angle $\theta$ with respect to the direction of a uniform
constant magnetic field ${\bf B}$, for $B=0,\, 0.1,\, 0.175$ and $0.2\,$a.u.
Three inclinations $\theta=0^\circ,\,45^\circ,\,90^\circ$
are studied in detail with emphasis to the ground state $1_g$.
Diamagnetic and paramagnetic susceptibilities are calculated (for $\theta=45^\circ$ for
the first time), they are in agreement with the experimental data and with other
calculations. For $B=0,\, 0.1$ and $0.2\,$a.u. potential energy curves $E$ vs
$R$  are built for each inclination, they are interpolated by
simple, two-point Pad\'e approximant $Pade[2/6](R)$ with accuracy of not less than 4
significant digits.
Spectra of rovibrational states are calculated for the first time.
It was found that the optimal configuration of the ground state for $B \leq
B_{cr}=0.178\,$a.u. corresponds always to the parallel configuration, $\theta=0$, 
thus, it is a $^1\Si_g$ state. The state $1_g$ remains bound for any magnetic field,
becoming metastable for $B > B_{cr}$, while for $B_{cr} < B < 12$\,a.u. the ground 
state corresponds to two isolated hydrogen atoms with parallel spins.
\end{abstract}


\end{frontmatter}


\section{Introduction}

More than fifty years have passed since it was predicted that extremely strong magnetic
fields up to $B=10^{14}-10^{16}\,$G  ($B\sim 4\times 10^{4-6}\,$a.u.), which are by far beyond
those that can be reached in the laboratory, could exist
in neutron stars remnant of a supernova explosion
as effect of the magnetic field flux compression~\cite{Woltjer:1964}
(see also \cite{Pacini:1967,Gold:1968,Goldreich:1969}).
As for magnetized white dwarfs, the surface magnetic field can reach $B \sim
10^{9}\,$G  (see e.g. \cite{Garcia:2016} and references therein). Soon
afterwards it was recognized that the structure of atoms and molecules might be
qualitatively different under strong magnetic fields $B\gtrsim B_0$
($B_0=1\,{\rm a.u.} \equiv 2.35\times 10^9\, {\rm G} = 2.35\times 10^5\,{\rm T}$)~\cite{Ruderman:1971,Kadomtsev:1971a,Kadomtsev:1972}
from the field-free case. The electronic clouds assume a well-pronounced cigar-like form, and molecules
become oriented along the magnetic line. Eventually,
the problem becomes quasi-one-dimensional, where longitudinal and transverse
motions  of the electrons are almost separated. This gives hope to develop an analytical
theory in the domain of very strong magnetic fields. The situation gets much
more complicated in the domain of intermediate magnetic fields,
say, of order of $B \sim 10^{-1}$\,a.u.,
where quadratic corrections to the linear Zeeman effect become significant.
This domain is `slightly' above the magnetic fields reachable in the laboratory.
In this case we do not see hope to develop analytic approaches.
We will call the fields $0.01 \lesssim B \lesssim 1\,$a.u. the {\it
intermediate} magnetic fields.

Due to mainly technical difficulties in solving the Schr\"odinger equation in
the presence  of intermediate and strong magnetic fields, only a relatively small
number of simple atomic and molecular systems has been studied. Naturally, the
hydrogen atom ${\rm H}$ and the hydrogen molecular ion ${\rm H}_2^+$ are the
most studied systems, see e.g.~\cite{Kravchenko:1996} and
\cite{Turbiner:2006PR}, respectively, and references therein. The first
quantitative study of the ${\rm H}_2$ molecule was carried out by one of the
present authors in 1983~\cite{Turbiner:1983}.
In the majority of studies of molecules and molecular ions all non-adiabatic
terms in the Hamiltonian are neglected by assuming an infinite nuclear mass
(what is usually called Born-Oppenheimer (BO) approximation of zeroth order).
The fact is that in both the ${\rm H}$ and $\rm H_2^+$ systems the binding energy
grows dramatically with an increase of the magnetic field strength, it
hints at the possible existence of unusual chemical species in strong magnetic
fields.
Other simple, traditional such as $\rm H_3^+$~\cite{Turbiner:2007PRA,MED13:9871}, and exotic compounds  mainly formed by protons and/or
$\alpha$-particles (helium nuclei) and one or two electrons have been studied to a
certain degree. For a discussion, see~\cite{Turbiner:2006PR} for one-electron
systems and~\cite{Turbiner:2010} for two-electron systems.

Recently, a detailed study of the H$_2^+$ molecular ion in inclined
configuration  (when the molecular axis and the magnetic line form some non-zero angle)
was carried out for intermediate and strong magnetic fields~\cite{Turbiner:2006PR,Medel:2015}.
It was shown that for the ground state the optimal configuration is always
parallel,  where the molecular axis and magnetic field direction coincide. The
spectra of rovibrational states was exhaustively studied.

As for the ${\rm H}_2$ molecule, it was found long ago that the minimal energy (ground) state
evolves with magnetic field strength being realized by different states
depending  on the strength of the magnetic field, see~\cite{Detmer:1997,Detmer:1998a}
and references therein.
At zero and weak magnetic fields, the ${\rm H}_2$ ground state is realized by
the  spin-singlet $S=0$, ${}^1\Si_g$ state in parallel configuration, but with
the magnetic field strength increasing to above the critical field strength of
$B_{cr} = 0.178\,$a.u., see below, the ground state changes to a spin-triplet $S=1$,
${}^3\Si_u$, state which is a repulsive state (!). It corresponds to two hydrogen atoms
at large distances with electron spins antiparallel to the magnetic field,
hence, the hydrogen molecule does not exist as a compact system.
It is worth noting that this value of the critical magnetic field was
calculated accurately in present paper and confirms the rough estimate
$B_{cr}\simeq 0.2\,$a.u. from~\cite{Detmer:1997}.
Nonetheless, for even stronger magnetic fields, $B\gtrsim 12\,$a.u., the
ground state is realized by a spin triplet $S=1$, $^3\Pi_u$ state,
see~\cite{Detmer:1998a} and references therein.
A similar behavior is observed in the case of the linear ${\rm H}_3^+$ molecular
 ion in strong magnetic fields: the ground state evolves from the spin-singlet
$^1\Si_g$ state for weak magnetic fields $B\lesssim 5\times 10^8\,$G ($\simeq
0.2\,$a.u.) to a weakly bound spin-triplet $^3\Si_u$ state for intermediate
and strong fields and, eventually,
to a spin-triplet $^3\Pi_u$ state for magnetic fields $B\gtrsim 5\times
10^{10}\,$G  ($\simeq 21\,$a.u.)~\cite{Turbiner:2007PRA}.  In such studies
the parallel configuration of the molecular axis and the magnetic field
direction is explicitly assumed.  Non-aligned configurations, where the
molecular axis is not parallel to the magnetic field direction, have received
much less attention. This is due to the fact that such
configurations require a much larger computational effort to reach the
accuracies obtained in the parallel case. The present authors are not aware of
any studies of inclined configurations for the H$_2$ molecule for $B \lesssim
0.2\,$a.u.

The goal of this paper is to study the hydrogen molecule \hh arbitrarily oriented
i.e. with the molecular axis forming an angle $\theta$
with respect to the direction of a uniform magnetic field ${\bf B}$ in lowest
spin-singlet state $1_g$.
The magnetic field strengths of interest in this work are chosen to be
$B=0,\, 0.1,\, 0.175$ and $0.2\,$a.u. (equivalently, $0, 2.35\times
10^8,  4.1\times 10^8$ and $4.7\times 10^8\,$G), where the ground state is
realized by the spin-singlet state $^1\Si_g$ at $\theta=0$ for $B < B_{cr}$. We
use the variational method with trial functions designed following a criterion
of physical adequacy \cite{Turbiner:1984,Turbiner:2006PR}. Three inclinations
$\theta=0^\circ,\,45^\circ$ and $90^\circ$ will be studied in detail and
the potential energy curves for each inclination and each magnetic field will
be constructed. The two-dimensional potential
energy surfaces are obtained by interpolation in the $\theta$ coordinate.
This allows us to calculate for the first time the lowest rovibrational levels of
the \hh molecule in weak and intermediate magnetic fields, where this molecule
exists as a compact object. A study of the magnetic susceptibility of the \hh
molecule is also performed. We will follow in presentation our previous work
on ${\rm H}_2^+$ in weak and intermediate magnetic fields~\cite{Medel:2015}.
Atomic units will be used through the text.

\section{The Hamiltonian and generalities}

We consider the hydrogen molecule ${\rm H}_2$ interacting with an external
magnetic field ${\bf B}$.
The origin of coordinates is chosen in the
midpoint of the line connecting the nuclei (molecular axis). The molecular axis
in turn forms an angle $\theta$ with respect to the magnetic field direction
(chosen to coincide with the $z$-axis). A convenient gauge which describes a
magnetic field oriented parallel to  the $z$-axis, is the linear gauge
\begin{equation}
\label{lgauge}
\hat{A}=B[(\xi-1)y,\xi x,0],
\end{equation}
where $\xi$ is a parameter.
If $\xi=0$ the linear gauge is reduced to the Landau gauge, and if $\xi=1/2$
then the symmetric gauge is obtained. In approximate variational calculations
the parameter $\xi$ is considered as an extra variational parameter.

Since the nucleus mass is by far larger than the electron mass, we can neglect
all non-adiabatic coupling terms in the Hamiltonian to obtain  the order zero
BO approximation. Thus, the electronic Hamiltonian in atomic units
($\hbar=m_e=c=1$) is given by
\begin{eqnarray}
 \label{Ham}
\hat{H}_e&=&
-\frac{1}{2}\sum_{i=1}^2\nabla_i^2-iB\sum_{i=1}^2\left(
(\xi-1)y_i\pa_{x_i}+\xi x_i\pa_{y_i} \right)
 +  {\mathbf S}\cdot {\mathbf B}
\nonumber \\
 & & +\frac{1}{2}B^2\sum_{i=1}^2\left( \xi^2 x_i^2+(\xi -1)^2y_i^2
 \right)-\sum_{i=1}^2\left(\frac{1}{r_{ia}}+\frac{1}{r_{ib}}\right)
 +\frac{1}{r_{12}}+\frac{1}{R}\,,
\end{eqnarray}
where $\nabla_i$ is the Laplacian operator
with respect to the coordinates of the $i$-th electron ${\bf r}=(x_i,y_i,z_i)$,
$r_{ia,ib}$ are the distances between the $i$-th electron and the nuclei $a$ or
$b$, respectively, $r_{ij}$ is the distance between the electrons and $R$ is
the distance between the nuclei. As usual, the contribution to the energy due to
the Coulomb interaction between the nuclei ($1/R$) is treated
classically.
Hence, $R$ is considered an external parameter. In the particular case
$\theta=0^\circ$, the component of the
angular momentum along the $z$-axis is conserved and the term linear in $B$
becomes $\frac{1}{2}{\mathbf L}\cdot {\mathbf B}$ for $\xi=\frac{1}{2}$. The
spin Zeeman term ${\mathbf S}\cdot {\mathbf B}$ with the total electron spin
${\mathbf S}={\mathbf S}_1 + {\mathbf S}_2$ is included in the Hamiltonian.
However, for the spin-singlet states with  ${\mathbf S}=0$ this term does not
contribute to the total energy and can be excluded.

Finally, the nuclear motion can be treated as vibrations and rotations
following the BO approximation with the electronic energy
taken as the potential in the nuclear Hamiltonian.

\section{The trial function}

Following physical relevance arguments (see, e.g. \cite{Turbiner:1984}) we
designed a spatial trial function
which   is    a product of Landau orbitals, Coulomb orbitals and a correlation term in
exponential form:
\begin{equation}
 \label{trial}
 \psi({\bf r}_1,{\bf r}_2)=\prod_{k=1}^2 ( e^{
-\alpha_{ka}r_{ka}-\alpha_{kb}r_{kb}-\frac{B\beta_{kx}}{4} x_k^2
 -\frac{B \beta_{ky}}{4} y_k^2})e^{\alpha_{12}r_{12}}
\end{equation}
where $\alpha_{ka,kb}$, $\beta_{kx}$, $\beta_{ky}$ with $k=1,2$ as well as
$\alpha_{12}$ are variational parameters. In (\ref{trial}) the variational
parameters $\al_{k{a}}, \al_{k{b}}$ ($k=1,2$) have the meaning of screening (or
anti-screening) factors (charges) for the nucleus ${a,b}$ respectively,
as it is seen from the $k$-th electron. The variational parameters
$\beta_{k{x}}$,  $\beta_{k{y}}$ account for the  screening (or anti-screening)
factors for the magnetic field seen from  $k$-th electron in $x, y$ direction
respectively, and the  parameter   $\al_{12}$ ``measures" the screening (or
anti-screening) of the electron correlation interaction.
This spatial function reproduces adequately the behavior of the electrons
near the Coulomb singularities and the harmonic oscillator at long distances
arising from the magnetic field.
In a certain way the trial function (\ref{trial}) is a generalization of the
trial function presented in~\cite{Turbiner:2007}
for the field free case. It reproduces two physical situations:
for small internuclear distances the trial function (\ref{trial}) mimics the
interaction ${\rm H}_2^++e$
(if $\alpha_{1a}=\alpha_{1b}$ and $\alpha_{2a}=\alpha_{2b}$)
while for large internuclear distances it mimics the interaction
${\rm H}-{\rm H}$ (if $\alpha_{1a}=\alpha_{2b}$ and $\alpha_{1b}=\alpha_{2a}$).

We consider a trial function which is a superposition of three Ans\"atze:
a general Ansatz of the type (\ref{trial}), a ${\rm H}-{\rm H}$ type Ansatz and a
${\rm H}_2^++e$ type Ansatz
\begin{equation}
 \label{superposition}
 \Psi=A_1\psi + A_2 \psi_{H+H}+A_3 \psi_{H_2^++e}\,,
\end{equation}
where $A_{1,2,3}$ are linear variational parameters.
Each Ansatz has its own set of variational parameters.
Without loss of generality $A_1$ may be set equal to the unity, therefore
the total number of variational parameters is $27$
including the internuclear distance $R$ and $\xi$ as variational parameters.

In the singlet state $(S=0)$ the trial function (\ref{superposition})
must be symmetric with respect to the exchange of the electrons and in the
gerade (g) state the trial function (\ref{superposition}) must be
symmetric with respect to the exchange of nuclei.
Therefore the operator
\begin{equation}
 \label{sym}
 (1+\hat{P}_{ab})(1+\hat{P}_{12})\,,
\end{equation}
where $\hat{P}_{ab}$ is the operator of symmetrization of nuclei and
$\hat{P}_{12}$ is the operator of symmetrization of the electrons,
must be applied to the trial function (\ref{superposition}).

The calculation of the variational energy using the trial function (\ref{trial})
 involves two major parts:
(i)  6-dimensional numerical integrations which were implemented by an adaptive
multidimensional integration $C$-language routine ({\sl
cubature})~\cite{GenzMalik1980}, and (ii) a minimizer which  was implemented
with the Fortran minimization package MINUIT from CERN-LIB. Our $C$-Fortran
hybrid program was parallelized using MPI.
The 6-dimensional integrations were carried out using a dynamical partitioning
procedure: the domain of integration is manually divided into sub-domains
following the profile of the integrand. Then each sub-domain is integrated on
separated processors using the routine {\sl CUBATURE}. In total, we have a
division into $960$ subregions for the numerator and $\sim 1000$ for the
denominator of the variational energy. With a maximal number of sampling points
$\sim 10^8$ for the numerical integrations for each subregion, the time needed
for one evaluation of the variational energy (two integrations) is $2 \times
10^3\,$ seconds ($\sim 37\,$min) with 96 processors at the cluster {\it KAREN}
(ICN-UNAM, Mexico). It was checked that this procedure stabilizes the estimated
accuracy and is reliable in the first three to four decimal digits. However, in
order to localize the domain of minimal parameters, a minimization
procedure with much less sample points was used in each sub-domain,
and a single
evaluation of the energy usually took $\sim 15-20\,$mins. Once a domain is
roughly  localized, the number of sample points is increased by a factor of $\sim
10^2$. Typically, a minimization process required several hundreds of
evaluations. As a general  strategy, the variational energy corresponding to the
general Ansatz only is calculated in first place. Then, either the  ${\rm
H}-{\rm H}$ type Ansatz or the ${\rm H}_2^++e$ type Ansatz is added as a first
correction, depending on which configuration yields a better variational result,
and the energy is minimized using the superposition of two Ans\"atze.  Eventually,
the remaining configuration is included in the final trial function and
a final minimization is carried out. The whole process is very lengthy and
cumbersome due to the absence of a fast minimization procedure. Computations were
mainly performed in parallel on 96 processors on the  cluster {\it ROMEO} at the University
of Reims, France, and on the cluster {\it KAREN}  at ICN-UNAM, Mexico.

\section{Results}

\begin{figure}[h!]
\begin{center}
 \includegraphics[angle=-90,width=120mm]{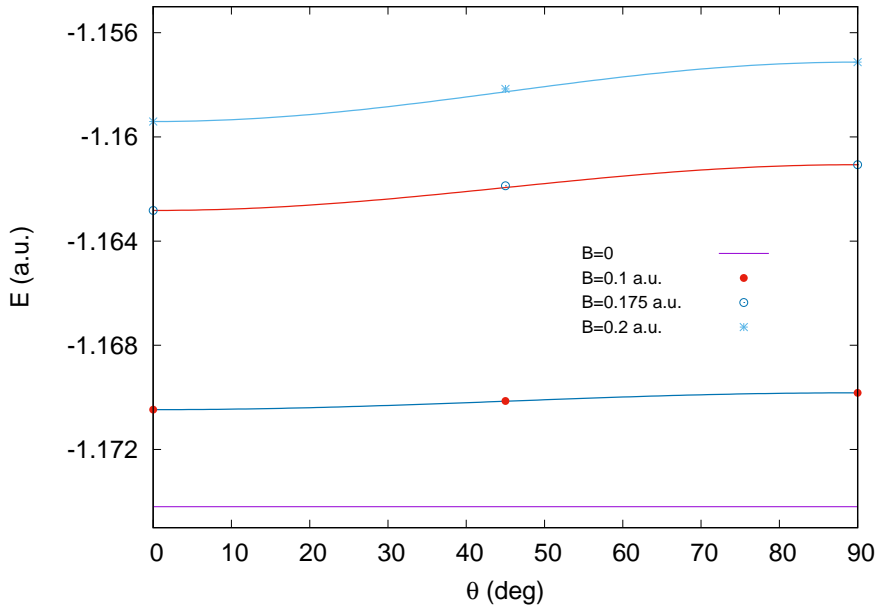}
 \caption{\label{H2EvsTheta}
  Total energy $E$ of ${\rm H}_2, 1_g$ state {\it vs} inclination $\theta$ for
  $B=0$ and $B=0.1, 0.175$, and $ 0.2 \,$a.u. The solid lines correspond to
  the hindered rotator model $E(B,\theta) =  E(B,0^\circ) + A \sin^2(\theta)$,
  where $A= (E(B,90^\circ)-E(B,0^\circ))$,     see Eq. (\ref{pot}).}
\end{center}
\end{figure}

\begin{figure}[h!]
\begin{center}
 \includegraphics[angle=-90,width=120mm]{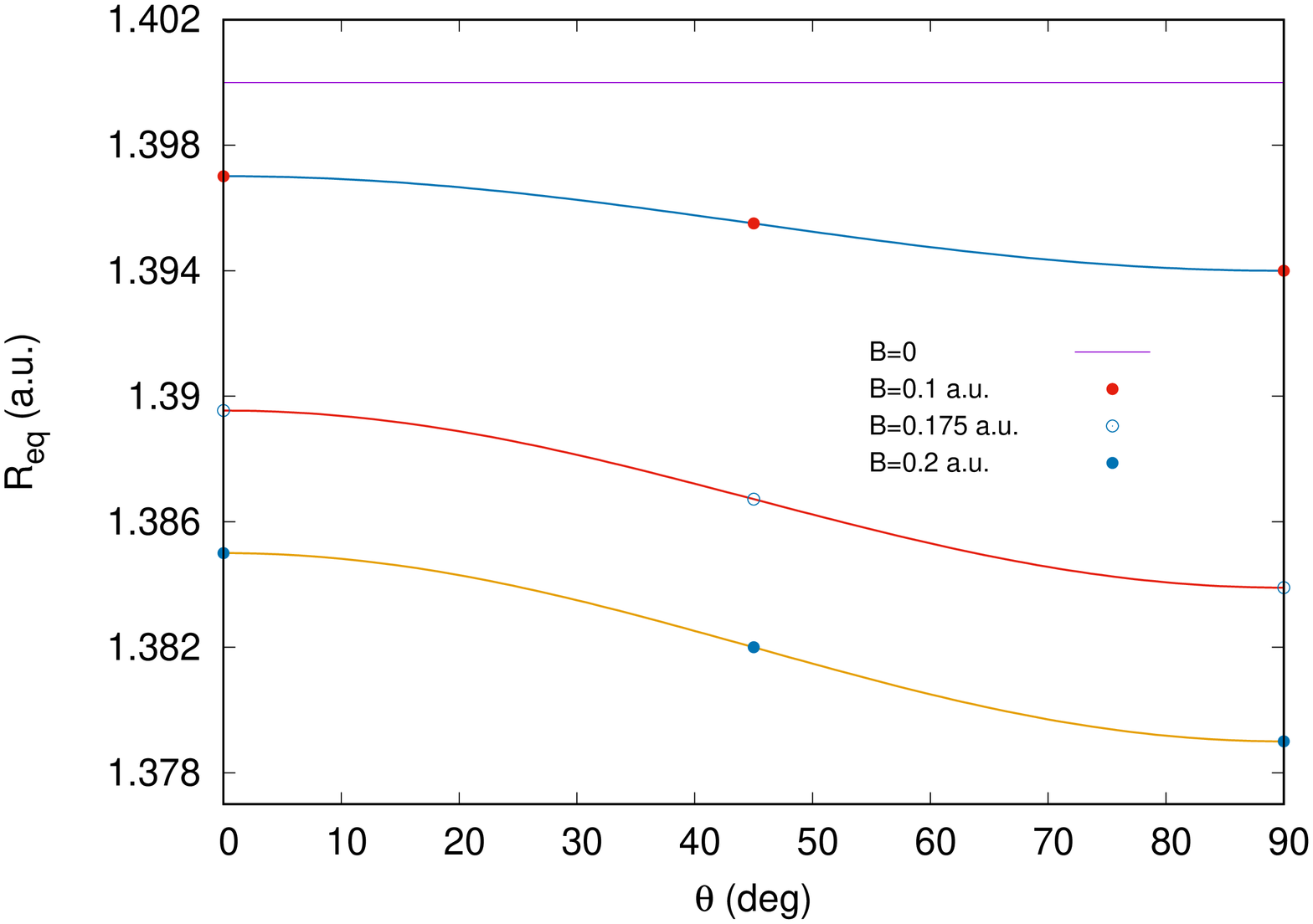}
 \caption{\label{H2RvsTheta}
  Equilibrium distance $R_{eq}$ of $\rm H_2$ in the $1_g$ state {\it vs} inclination angle
  $\theta$ for $B=0, 0.1,0.175, 0.2\,$a.u. The solid lines correspond to
  $R_{eq}(B,\theta) =  R_{eq}(B,0^\circ) + C \sin^2(\theta)$,
  where $C = R_{eq}(B,90^\circ)-R_{eq}(B,0^\circ)$ (c.f. Eq
  (\ref{pot})).}
\end{center}
\end{figure}

The electronic energies and the equilibrium distances of ${\rm H}_2$ in the
$1_g$  state are presented in Table~\ref{tablaresultados} for magnetic fields
$B=0,\,0.1,\, 0.175\,$  and $0.2\,$a.u. Variational energies
indicate that for $B\leq B_{cr}=0.178\,$a.u. the lowest energy state of
${\rm H}_2$ is realized by the $1_g$ state in parallel configuration.
For $B_{cr}=0.178\,$a.u. the energy of the $1_g$ state at the equilibrium
minimum  coincides with the energy of two hydrogen atoms infinitely separated
and having both electron spins antiparallel to the magnetic field direction.
Thus, for $B=0.2\,$a.u. the state $1_g$ is, in fact, a m-eta-stable state.
We studied the geometrical configurations with angles $\theta=0^\circ,45^\circ$
and $90^\circ$ between the magnetic field direction and molecular axis in
great detail, while some sample calculations were carried out for the intermediate angles $\theta=15^\circ, 30^\circ$
and $60^\circ, 75^\circ$ to check the  smoothness of the angular
dependence of both, the energy and the equilibrium distance (see
below). For all inclinations the potential energy curve $E$ vs $R$ exhibits a
well pronounced minimum at a finite internuclear distance $R$. As the magnetic
field increases, for any given inclination the system becomes more strongly bound (the
binding energy increases) and more compact (the internuclear equilibrium
distance $R_{eq}$ is reduced), see Table~\ref{tablaresultados}. Note that for the field-free
case $B=0$ our energy is in agreement with one of the most
accurate results~\cite{Sims:2006} in $\sim 2 \times 10^{-4}$\,a.u. in spite of
the very simple form of the trial function that we used. We must emphasize that for
parallel configuration $\theta=0$ our energies are systematically better than
the ones from~\cite{Detmer:1997} in 3 decimal digits (d.d.), which leads to a more accurate value of
the critical magnetic field strength $B_{cr}$.
For a given magnetic field, the total energy increases while the
equilibrium distance $R_{eq}$ shows a small decrease with growth of the inclination
angle from $\theta=0$ to $90^\circ$, see Figures~\ref{H2EvsTheta}
and~\ref{H2RvsTheta}. Such an increase in $E$, and decrease in $R_{eq}$, are
more pronounced as the magnetic field increases. Thus, for all magnetic fields
studied, the optimal configuration  corresponds to the parallel configuration as
it is expected.
The angular dependence of the variational energy $E(B,\theta)$ and
the equilibrium distance $R_{eq}(B,\theta)$ for a fixed
magnetic field strength $B$ is very simple and is well-described by the
hindered rotator model, see Eq. (\ref{pot}) and captions of
Figs.~\ref{H2EvsTheta} and~\ref{H2RvsTheta}. This observation is in agreement with
the test calculations for angles $\theta=15^\circ,
30^\circ, 60^\circ$ and $75^\circ$.
\begin{table}[t!]
\caption{
Total electronic energy and equilibrium distance of \hh in the state $1_g$ {\it
vs} magnetic field $B$ and inclination $\theta$ based on
trial function (\ref{superposition}), see text. Energies $E$ and
equilibrium distances $R_{eq}$ rounded to 5th and 3rd d.d., respectively.
${}^*$ For $B=0.2$\,a.u. the $1_g$ state is no longer the ground state.
Results marked $\dagger$ are from Ref.~\cite{Detmer:1997}, those marked $\ddagger$
from~\cite{Sims:2006} (rounded).
The binding energy $E_{\rm bind}\equiv 2E({\rm H}) - E({\rm H}_2)$ with
respect to dissociation to ${\rm H} +{\rm H}$ is shown in the last column,
where the energies for the ${\rm H}$ atom in ground state are taken from~\cite{Kravchenko:1996}.
}
 \label{tablaresultados}
\centering
\begin{tabular}{c|cccc}
\hline \hline
$B\,$(a.u.) &  $\theta\,$(degrees) & $E\,$(a.u.)             & $R_{eq}\,$(a.u.)
& $E_{\rm bind}\,$(a.u.)  \\
\hline
$0.0$       & -                    & $-1.17420$              & $1.40$
&  $0.17420$  \\
            &                      & $-1.174476^{\ddagger}$  & $1.40$           &  \\
\hline
$0.1$       & $0$                  & $-1.17047$              & $1.397$          &  $0.17542$\\
            &                      & $-1.16965^{\dagger}$    & $1.39^\dagger$   &  \\
            & $45$                 & $-1.17014$              & $1.396$          &  $0.17508$\\
            & $90$                 & $-1.16983$              & $1.394$          &  $0.17477$    \\
\hline
$0.175$     & $0$                  & $-1.16282$              & $1.390$          &  $0.17768$    \\
            & $45$                 & $-1.16187$              & $1.387$          &  $0.17673$    \\
            & $90$                 & $-1.16107$              & $1.384$          &  $0.17592$    \\
\hline
$0.2^*$     & $0$                  & $-1.15941$              & $1.385$          &  $0.17864$    \\
            &                      & $-1.15877^{\dagger}$    & $1.39^\dagger$   & \\
            & $45$                 & $-1.15816$              & $1.382$          &  $0.17740$    \\
            & $90$                 & $-1.15713$              &  $1.379$         &  $0.17636$    \\
\hline \hline
\end{tabular}
\end{table}

\section{Potential Energy Curves}

Potential energy curves $E$ vs $R$ of the  state $1_g$ of the ${\rm H}_2$
molecule in
magnetic fields \hbox{$B=0, 0.1, 0.2\,$a.u.} and inclinations
$\theta=0,45^\circ, 90^\circ$ are built from variational results obtained in
the domain $R \in [1,2]\,$a.u.
and extended beyond following the procedure discussed in~\cite{Turbiner:2018}
for approximating potential curves in diatomic molecules (see also references
therein).
It is evident that the asymptotic behavior of the electronic
energy of ${\rm H}_2$ at small distances $R\rightarrow 0$ is given by
\begin{equation}
\label{EvsRa1}
E\  \approx \ \frac{1}{R}\ +\ E_{ {\rm He}}(B)\ +\ c_1 R\ +\ O(R^2)\ ,
\end{equation}
where $E_{{\rm He}}(B)$ is the ground state energy of the helium atom in a
magnetic
field $(B)$ (the so-called united atom limit), and the coefficient in front of
$R$ depends on the magnetic field and the inclination $\theta$,
$c_1=c_1(B,\theta)$; at $B=0$ this coefficient vanishes $c_1=0$ (see
\cite{Turbiner:2018} and references therein).
As for the asymptotic limit $R \rar \infty$, the expansion of the energy $E$ is given by
\begin{equation}
\label{EvsRa2}
 E\approx E_{2{\rm H}}(B)\ +\ \frac{c_5}{R^5}\ - \ \frac{c_6}{R^6} +\
 \frac{c_7}{R^7}\
 +\ O\left(\frac{1}{R^8}\right)\ ,
\end{equation}
where $E_{2{\rm H}}(B)$ is the energy of two (infinitely separated) hydrogen
atoms in their ground state in the magnetic field of strength $B$  (however,
with opposite electron spin projections so that ${\mathbf S}\cdot {\mathbf
B}=0$), the term  $\propto 1/R^5$  corresponds to the quadrupole-quadrupole
interaction (repulsive for $0, 90^{\circ}$ and attractive for $45^{\circ}$)
between two separated hydrogen atoms in the magnetic field (which is the leading
order interaction at $R \to \infty$). The term  $\propto 1/R^6$ corresponds to
the induced dipole-dipole interaction (in second order perturbation theory in
$1/R$ for $B=0$) between two separated hydrogen atoms (see \cite{Turbiner:1983}
and \cite{Kadomtsev:1972}). The  coefficients $c_{5,6,7}$  can depend on the
magnetic field strength and inclination $c_{5,6,7}=c_{5,6,7}(B,\theta)$.  In
absence of a magnetic field $c_{5,7}=0$. In general, the quadrupole-quadrupole
interaction energy (in a.u.) is given by
\begin{equation}
E_Q = \frac{3}{4} \frac{Q_{zz}^2(B)\, P_4(\cos \tha)}{R^5}\,,
\end{equation}
where $Q_{zz}$ is the quadrupole moment of the hydrogen atom in a magnetic field
of strength $B$ (see \cite{Turbiner:1983}), $P_4$ is 4th Legendre polynomial.
Thus,  the coefficient $c_5$ is known. For weak magnetic fields $B$ we use the
approximation the quadrupole moment in perturbation theory (see
\cite{Potekhin:2001})
\begin{equation}
Q_{zz} =  -\frac{5}{2} B^2 + \frac{615}{32} B^4 + \ldots\,.
\end{equation}

\begin{figure}[h!]
\begin{center}
\includegraphics[angle=-90,width=120mm]{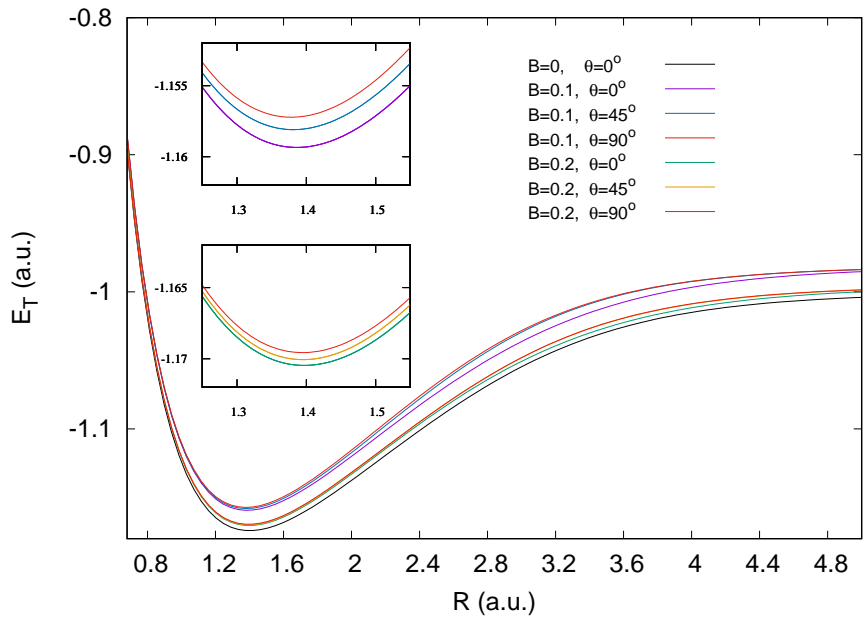}
\caption{\label{H2}
 Potential Energy curves of the $1_g$ ground state
 of the ${\rm H}_2$ molecule for $B=0,0.1,0.2\,$a.u. and
 $\theta=0,45^\circ, 90^\circ$. The insets show amplified
 energy curves for $B=0.1\,$a.u. (bottom), and $B=0.2\,$a.u. (top)
 around the equilibrium distance.
 For each given magnetic field, the lowest energy curve around the minimum
 always corresponds to the parallel configuration $\theta=0^\circ$, while the
 highest energy curve corresponds to the perpendicular configuration
 $\theta=90^\circ$.  For $B=0.2\,$a.u. the minimal energy of ${\rm H}_2$
 corresponds to the asymptotic energy of the repulsive $3_u$ triplet state (the
 energy of two hydrogen atoms infinitely separated with spins antiparallel to
 the magnetic field direction)  and lies below the minimum for the $1_g$ state.
 }
\end{center}
\end{figure}

Now we interpolate both asymptotic expansions (\ref{EvsRa1}) and (\ref{EvsRa2})
via the two-point Pad\'e approximant $Pade[N/N + 4](R)$ with $N=2$ as the
minimal degree, which guarantees that the expansions (\ref{EvsRa1}) and
(\ref{EvsRa2}) are described functionally correct,
\begin{equation}
\label{fitPotCur}
 E(R)\ =\ \frac{1}{R}\ \frac{a_0 +a_1 R + a_2 R^2 }{ \left( b_0 + b_1 R +b_2
    R^2  + b_3 R^3 + b_4 R^4 + b_5 R^5 + b_6 R^6\right)}+E_{2{\rm H}}(B)   \ ,
\end{equation}
where the constraints
 $$b_0=a_0\ ,\ b_1=a_0 \left( E_{2{\rm H}}(B) -  E_{ {\rm He}}(B)\right) +
  a_1\ ,$$
are imposed in order to reproduce the first two leading terms in (\ref{EvsRa1})
exactly plus the condition $c_5=\frac{3}{4} Q_{zz}^2(B)$, it implies  the
relation $$a_2 = c_5 {b_6}\ .$$ Without loss of generality we can set $a_0=1$.
Therefore, we have six free parameters $a_1$, $b_2$, $b_3$, $b_4$, $b_5$, $b_6$
to fit the variational energies at internuclear distances near the equilibrium,
$R\in [1,2]\,$a.u. for  $B=0, 0.1, 0.2\,$a.u. and inclinations
$\theta=0,45^\circ,90^\circ$.
The value of the parameters is presented in Table~\ref{tablaS4}.
The potential energy curves are shown in Fig~\ref{H2}.
In general, the curves (\ref{fitPotCur}) reproduce four d.d. in energy at
$R\in [1,2]\,$a.u.

\begin{table}[t!]
\caption{\label{tablaS4}
 Fitted parameters (rounded to 5 d.d.) in the Pad\'e approximant (\ref{fitPotCur}) for
 the \hh potential energy curves \hbox{$E$ vs $R$} for $B=0, 0.1, 0.2\,$a.u., see Fig.~\ref{H2}.
}
\centering
\begin{tabular}{cccccccc}
\hline \hline
$B$(a.u.)      & $\theta$    & $a_1$      & $b_2$      & $b_3$       & $b_4$    & $b_5$      &  $b_6$   \\ \hline
 $0$            & $0^\circ$  &  -1.28814  & 1.09317    & -0.94343    &  0.79009 &  -0.28161  &  0.04758 \\
 $0.1$          & $0^\circ$  &  -1.29188  & 0.97020    & -0.64470    & 0.51708  & -0.17165   & 0.03136  \\
                & $45^\circ$ &  -1.29133  & 1.16648    & -1.12137    & 0.94652  & -0.34039   & 0.05644  \\
                & $90^\circ$ &  -1.29330  & 1.30768    & -1.38159    & 1.13784  & -0.40315   & 0.06395  \\
 $0.2$          & $0^\circ$  &  -1.30393  & 1.04298    & -0.80653    & 0.65880  & -0.22496   & 0.03886  \\
                & $45^\circ$ &  -1.30284  & 1.09918    & -0.99912    & 0.90103  & -0.35196   & 0.06408  \\
                & $90^\circ$ &  -1.30207  & 1.13486    & -1.06132    & 0.93558  & -0.35573   & 0.06400  \\
\hline \hline
\end{tabular}
 \end{table}

\section{Magnetic Susceptibility}

The trial function (\ref{superposition}), in spite of its simplicity, incorporates
accurately the many physics features of the \hh molecule in a magnetic field.
In order to verify this assertion for weak magnetic fields we calculated the
magnetic susceptibility (magnetizability). To make this calculation
we follow the recipe proposed in our work on ${\rm H}_2^+$ \cite{Medel:2015}.

It is well known that the response of the molecule to an external magnetic
field falls into two parts: diamagnetic and paramagnetic.
Correspondingly, there are two contributions to the magnetic susceptibility:
a paramagnetic, $\chi^p$, originating from the linear Zeeman
term in the Hamiltonian (\ref{Ham}) when treated within second order perturbation
theory in powers $B$, and a diamagnetic, $\chi^d$, coming from the quadratic
Zeeman term $\sim B^2$ in the first order of  perturbation theory. Thus, the total
magnetic susceptibility is the sum of the two terms $\chi=\chi^d+\chi^p$.

In general, the magnetic susceptibility tensor $\chi_{\al\be}$  is defined by the
coefficients in the operator
 \[
   {\cal H}' = -\frac{1}{2}  \sum_{\al,\beta}\chi_{\al\beta}B_\al B_\beta\ ,
 \]
with $B_\al, B_\beta$ being the components of the magnetic field.
As for the diamagnetic susceptibility it is given in first order PT in $B^2$ as
\begin{equation}
 \label{diatensor}
 \chi^{d}_{\alpha \beta}\ =\ -\frac{1}{4}\sum_{i=1}^2\left( \langle {\bf r}_i^2
 \rangle\delta_{\alpha\beta}\ -\ \langle r_{i,\alpha}
r_{i,\beta}\rangle\right)\ ,
\end{equation}
where both the expectation value $\langle {\bf r}^2_i\rangle$ of the position
vector squared of the $i$-th electron and the 2nd order tensor $\langle
r_{i,\alpha} r_{i,\beta}\rangle$, $\alpha,\beta  = x, y, z$ are taken with
respect to the field-free wavefunction at equilibrium distance $R_{eq}$.
If the magnetic field direction is chosen along the $z$-axis, ${\bf B}=B\hat{z}$
the tensor $\chi^d_{\alpha \beta}$ appears in diagonal form and contains a single
non-zero component, $\chi^d_{zz}\equiv \chi^d$,
\begin{equation}
 \label{diatensoruse}
 \chi^d=-\frac{1}{4}\sum_{i=1}^2(\langle x_i^2 \rangle +\langle y_i^2\rangle)\,,
\end{equation}
where the symmetric gauge is assumed to be taken. On the other hand,
the paramagnetic contribution to the susceptibility is much more difficult to
calculate, since it occurs in second order PT. In general, the paramagnetic
susceptibility is much smaller than the diamagnetic one. In principle, this
contribution to the susceptibility can be easily evaluated as the difference
$\chi^p =\chi -\chi^d$, where $\chi$ is the total magnetic susceptibility at a
given inclination. As for the ground state, the total magnetic susceptibility can
be calculated in a straightforward way as the coefficient in front of the $B^2$ term
in the energy expansion
\begin{equation}
\label{taylorms}
  E(B, \theta)=E(0) - \frac{1}{2} \chi(\theta) |{\mathbf B}|^2 + \ldots \ ,
\end{equation}
at $R=R_{eq}$.

The results for the susceptibilities are presented in Table~\ref{tablasuceptibilidad}
for inclinations $\theta=0^\circ, 45^\circ, 90^\circ$,
they are compared with the experimental data from Ramsey~\cite{Ramsey},
and with other calculations, when available. In general, all susceptibilities
grow with the inclination angle.
For $\theta=0$ our $\chi^d$ are larger than the values obtained in the past in
\cite{Riley:1977,KOLOS:1965}. They are closer to experimental data
being different from experimental data in one portion $\times 10^{-3}$ in
spite of the fact that our trial function is much simpler than the ones used in
\cite{Riley:1977} and~\cite{KOLOS:1965}. As for $\theta=45^\circ$, the susceptibilities
are calculated for the first time to the best of our knowledge, while for
$\theta=90^\circ$ our $\chi^d$ agrees in 2 d.d. with \cite{KOLOS:1965} and
differs from experimental data in $2 \times 10^{-2}$. Concerning $\chi^p$, it is
superior to all nine values calculated previously and collected in Table XII of
\cite{MARUANI}, however, it still differs from experimental data
in $\sim 20\%$. In general, our results for the susceptibility agree very well
with the experimental data and with other calculations.

\begin{table}[th!]
\caption{
Diamagnetic $\chi^d$, paramagnetic $\chi^p$ and total $\chi$ susceptibilities
of \hh in the state $1_g$ for different inclinations $\theta$ at $R=R_{eq}$.
The paramagnetic susceptibility $\chi^p$ obtained as the difference
$\chi^p=\chi - \chi^d$ (see text) is included for completeness.  The
expectation values  of the squared components of the position vector of each
electron $\langle x_{1,2}^2 \rangle$,  $\langle y_{1,2}^2 \rangle$  and
$\langle z_{1,2}^2\rangle$  (in a.u.) are also included for $B=0$ at the
equilibrium distance $R_{eq}=1.40\,$a.u.,  they were obtained using the trial function
(\ref{superposition}). $^{\rm exp}$ Experimental results from~\cite{Ramsey},
see also Table I in~\cite{Pflug:1977}. Results marked as
 $^a$ are from~\cite{Riley:1977},
 $^b$ from~\cite{KOLOS:1965},
 $^c$ from~\cite{MARUANI}.
}
\label{tablasuceptibilidad}
\centering
\begin{tabular}{ c ccc ccc }
\hline \hline
$\theta\,$  & $\langle x_{1,2}^2 \rangle$ & $\langle y_{1,2}^2 \rangle$ & $\langle z_{1,2}^2\rangle$ & $\chi^d$    & $\chi^p$   & $\chi$       \\ \hline
$0^\circ$   & $0.76465$     & $0.76465$     & $1.00929$      & $-0.7647$          & $0.0$      & $-0.7647$    \\
            & $0.7608^a$    & $0.7608^a$    & \ $0.9730^a$   & $-0.7608^a$        & $0.0^a$    &              \\
            & $0.76169^b$   & $0.76169^b$   & \ $1.02297^b$  & $-0.7617^b$        & $0.0^b$    &              \\
            &               &               &                & $-0.766^{\rm exp}$ &            &              \\    \hline
$45^\circ$  & $0.88697$     & $0.76465$     & $0.88697$      & $-0.8258$          & $0.0240$   & $-0.8046$    \\    \hline
$90^\circ$  & $1.00929$     & $0.76465$     & $0.76465$      & $-0.8870$          & $0.0258$   & $-0.8612$    \\
            & $0.9730^a$    & $0.7608^a$    & $0.7608^a$     & $-0.8669^a$        &            &              \\
            & $1.02297^b$   & $0.76169^b$   & $0.76169^b$    & $-0.8923^b$        &            &              \\
            &               &               &                &                    & $(0.027-0.082)^c$     &     \\
            &               &               &                & $-0.913^{\rm exp}$ & \ $0.022^{\rm exp}$ &    \\
\hline \hline
\end{tabular}

\end{table}

\section{Rovibrational levels}

The lowest rovibrational states of \hh and \dd were calculated for the field
strengths $B = 0.1 \, B_0$ and $B = 0.2 \, B_0$, where $B_0=2.35\times10^9$ Gauss $=2.35\times10^5$~T, as described in \cite{Medel:2015}.
To keep the present paper self-contained, the method is briefly summarized below.
Starting point is the nuclear Hamiltonian expressed in spherical coordinates,
\begin{equation} \label{NucHamil2}
\hat{\cal H}_{nuc}=-\frac{2}{M_s}\frac{1}{R}\frac{\partial^2}{\partial R^2}R +\frac{2}{M_sR^2}\hat{L}_R^2-\frac{1}{M_s}B\hat{L}_z
+\frac{1}{8M_s}B^2R^2\sin^2{\theta}+\tilde{V}(R,\theta).
\end{equation}
Here, $M_s$ denotes the total mass of the nuclei, $\hat{L}_z$ is the projection of angular momentum along $z$-axis
and $\theta$ the angle between the molecular and the $z$-axis.
The two-dimensional potential, $\tilde{V}(R,\theta)$, is parametrized as a hindered rotator, where
only the lowest expansion term is maintained, to yield
\begin{eqnarray}\label{pot}
\tilde{V}(R,\theta) & = & \tilde{V}(R,0)
+ \sum_n \frac{V_{90,n}(R)}{2} \left[ 1 - \cos (2 n \theta) \right] \nonumber \\
& \approx & \tilde{V}(R,0) + V_{90}(R) \sin^2 \theta
\end{eqnarray}
%
$V_{90}(R)=\tilde{V}(R,90)-\tilde{V}(R,0)$ is the barrier height for
a given value of~$R$.


The rovibrational wave function can be expanded in terms of
vibrational and rotational basis functions as
\begin{equation}\label{rv_basis}
\Psi(R,\theta,\phi) = \sum_{v,L} c_{v,L} \frac{ \xi_{v}(R;\underline{\theta}^{\prime})}{R}
Y_L^M(\theta,\phi)
\end{equation}
where $\xi_{v}(R;\underline{\theta}^{\prime})$ are solutions of the vibrational part of Eq.~(\ref{NucHamil2})
at the reference orientation $\theta^{\prime}$, chosen as  $\theta^{\prime}=0$.
These are obtained numerically using the renormalized Numerov algorithm.
The $Y_L^M(\theta,\phi)$ in the above equation are spherical harmonics.

In this basis, the matrix elements of the Hamiltonian in Eq.~(\ref{NucHamil2}) are given by
\begin{eqnarray}\label{Hnuc_mat2}
\left\langle v^{\prime} L^{\prime} M | \hat{\cal H}_{nuc} | v L M \right\rangle  & = &
E_v \delta_{L^{\prime} L}\delta_{v^{\prime} v}
+ \frac{2}{M_s} \left\langle v^{\prime} | \frac{1}{R^2} | v \right\rangle L(L+1) \delta_{L^{\prime}L} \nonumber \\
& - & \frac{B M }{M_s} \delta_{L^{\prime} L}\delta_{v^{\prime} v}  \nonumber \\
& + & \left[\frac{B^2}{12M_s} \langle v^{\prime} | R^2 | v \rangle
+\frac{2}{3}\langle v^{\prime} | V_{90}(R) | v \rangle \right] \delta_{L^{\prime}L} \nonumber \\
& - &\left[\frac{B^2}{12M_s} \langle v^{\prime} | R^2 | v \rangle
+\frac{2}{3}\langle v^{\prime} | V_{90}(R) | v \rangle \right]  \nonumber \\
& \times & (-1)^{M}\sqrt{(2L'+1)(2L+1)}   \nonumber \\
& \times & \left(
\begin{array}{ccc}
L&2&L'\\
0&0&0
\end{array}
\right)
\left(
\begin{array}{ccc}
L&2&L'\\
M&0&-M
\end{array}
\right)
\end{eqnarray}
The terms in parentheses are Wigner $3j$-symbols. The matrix~Eq.~(\ref{Hnuc_mat2}) is diagonal in $M$ as expected, since $M$ is
an exact quantum number. $L$-functions are coupled in steps of 2, conserving $z$-parity, $\pi=(-1)^{L+M}$. Diagonalization of the Hamiltonian matrix, Eq.~(\ref{Hnuc_mat2}), yields the eigenvalues and eigenvectors of the rovibrational problem.

We have computed the lowest rovibrational states for \hh and \dd.
Allowed rovibrational states must obey the permutational symmetry of the two identical nuclei.
In the case of \hh, with two fermions, the symmetry of the vibrational and rotational parts of the rovibrational wavefunction must be opposite,
while in the case of \dd, with two bosons, it must be the same if we consider {\em ortho} nuclear spins. For a rovibrational state of given vibrational quantum number, $v$,
and projection of the angular momentum on the magnetic field axis, $M$, the $z$-parities are thus
\begin{equation} \label{zpar}
\pi = (-1)^{M+v+1} = \left\{
\begin{array}{rl}
-(-1)^M  & {\rm for} ~v~ {\rm even} \\
(-1)^M  & {\rm for} ~v~ {\rm odd}
\end{array}
\right.
\end{equation}
for \hh, and
\begin{equation} \label{Dzpar}
\pi = (-1)^{M+v} = \left\{
\begin{array}{rl}
(-1)^M  & {\rm for} ~v~ {\rm even} \\
-(-1)^M  & {\rm for} ~v~ {\rm odd}
\end{array}
\right.
\end{equation}
for \dd.

The results of our calculations for the lowest vibrational states, $v=0, 1, 2, 3$ and $M \le 5$ are presented in Tables~\ref{Table:rotv0}--\ref{Table:rotv3} for \hh and
in Tables~\ref{Table:Drotv0}--\ref{Table:Drotv3} for \dd, for the magnetic field strengths $B = 0.1 \, B_0$ and $B = 0.2 \, B_0$.
We note that at $B = 0.2 \, B_0$ the molecule is meta-stable. As in our previous work on $\rm H_2^+$, two models have been considered: the approximate model 1,
in which off-diagonal terms in $v$ are omitted when setting up the rovibrational matrix, Eq.~(\ref{Hnuc_mat2}),
and model 2, in which they are included. The closeness of the two sets of results demonstrate that a simple expansion,
with just one vibrational function, yields a good approximation of the final rovibrational wavefunction, at least
for the lowest vibrational states. Therefore, in the full expansion of model 2, the coefficients $c_{v,L}$ allow
easy identification of the vibrational quantum number of each computed eigenstate.

All states are located above the rotational barrier, which is at $E_{barrier} = -1.16972 \, E_h$ for $B = 0.1 \, B_0$ and
$E_{barrier} = -1.15713 \, E_h$ for $B = 0.2 \, B_0$, respectively, and  hence, $L$, which is an exact quantum number in the field-free case,
can still be considered a ``good'' quantum number. It is interesting to analyse the orientation with respect to the magnetic
field axis of the lowest rovibrational state. The lowest state of \hh, at the field strength of $B = 0.1 \, B_0$, is located $0.0101\, E_h$,
or $2222\, \rm cm^{-1}$, above the barrier. Yet only one of the basis functions of the expansion in Eq.~(\ref{rv_basis}) contributes
effectively to its eigenvector, with coefficient $c_{0,1}=0.997$. The eigenfunction of the lowest state is thus
$\Psi(R,\theta,\phi) \sim \left[ \xi_{v}(R;\underline{\theta}^{\prime}=0)/R \right] Y_1^0(\theta,\phi)
\sim \left[ \xi_{v}(R;\underline{\theta}^{\prime}=0)/R \right] \cos \theta$, which shows that molecule essentially vibrates in the
direction of the magnetic field.

\begin{figure}[htp!]
\centering
\includegraphics[angle=-90, width=0.8\textwidth]{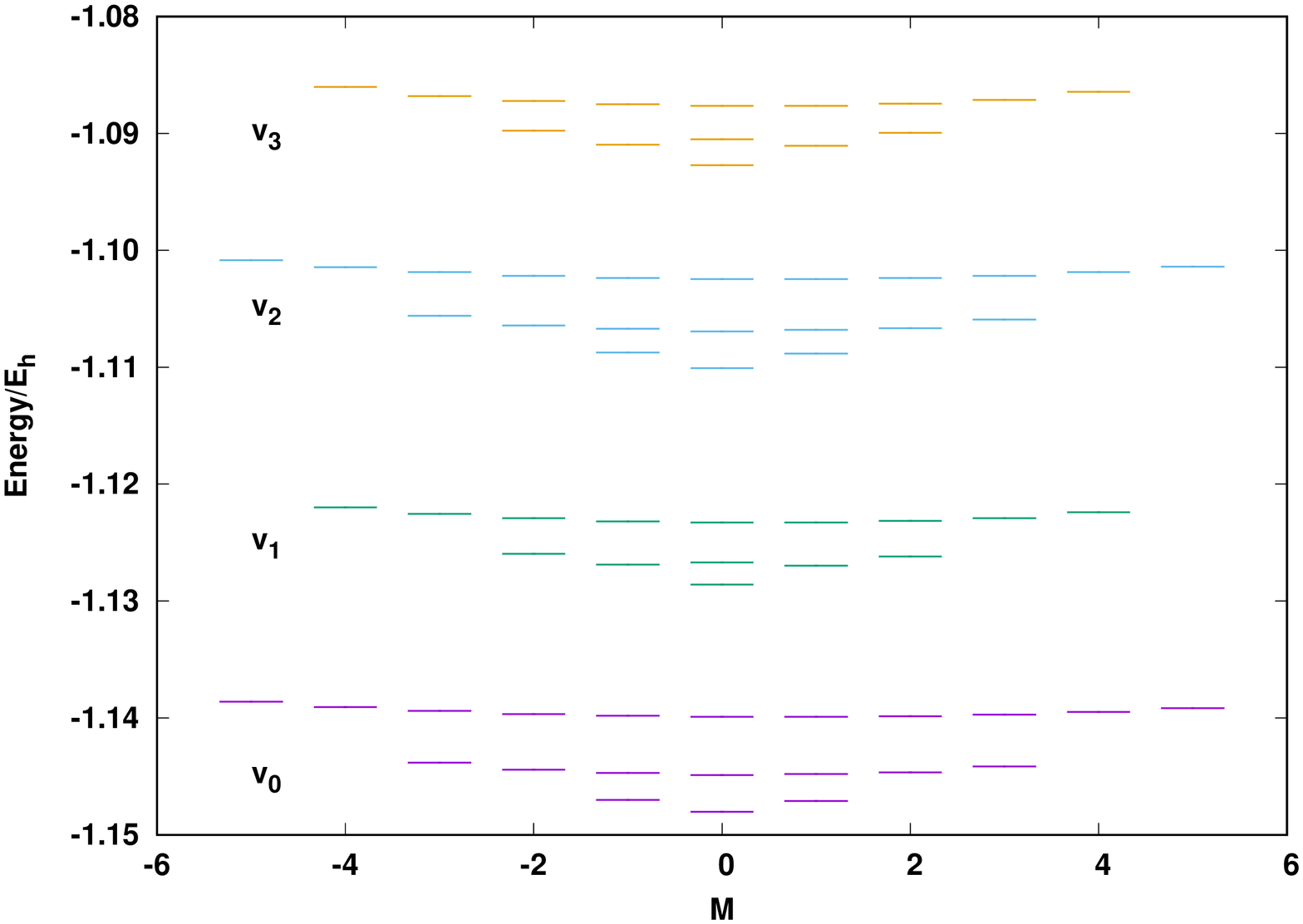}
\caption{Rotational structure, up to $L=5$, of the four lowest vibrational
states of \hh in the presence of an external magnetic field of $B=0.2\, B_0$.
}
\label{rovibplotB02}
\end{figure}

\begin{figure}[htp!]
\centering
\includegraphics[angle=-90, width=0.8\textwidth]{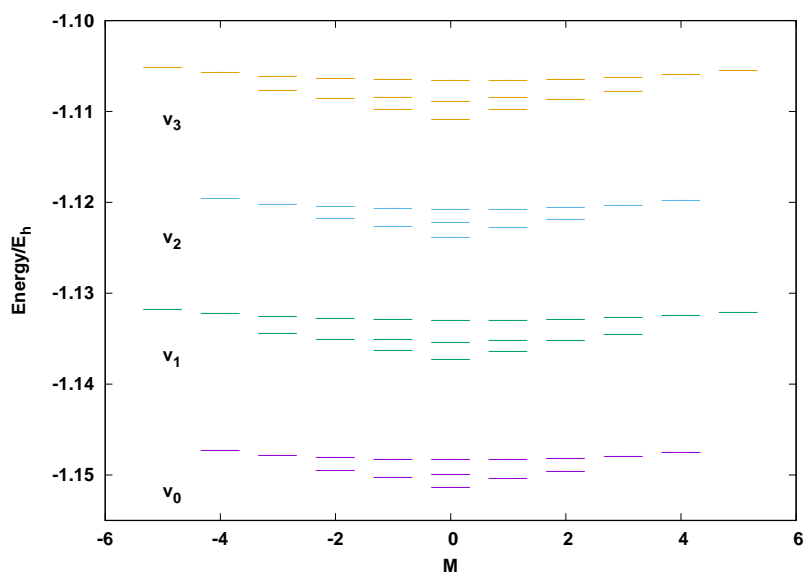}
\caption{Rotational structure, up to $L=5$, of the four lowest vibrational
states of \dd in the presence of an external magnetic field of $B=0.2\, B_0$.
}
\label{DrovibplotB02}
\end{figure}

In general, within each $L$-layer, the rotational energy of a vibrational state increases with $\left| M \right|$.
Figures~\ref{rovibplotB02} and ~\ref{DrovibplotB02} show some exceptions for the states $v=1,3$ of \hh and $v=0,2$ of \dd, where the $M=0$ state corresponding to $L=2$ is above $\left| M \right| = 1$. A similar effect has been observed in the case of $\rm H_2^+$ and $\rm D_2^+$. It is due to strong coupling of the $L=0$ and $L=2$
basis functions, a kind of Fermi resonance of the zero-order states with well-defined $L$. The effect scales as $B^2$ and is not visible for the lower field strength, $B = 0.1 \, B_0$. No strong effect can be seen for
the states  $v=0,2$ of \hh and $v=1,3$ of \dd, which have $L=1,3 \dots$, where the $L=1$ and $L=3$ layers are sufficiently separated in energy.

\clearpage

\begin{table}
\caption{
Rotational energy levels of \hh in presence of a uniform magnetic field $B$ for the vibrational state $v=0$. The pure
vibrational state ($L=0$ in the field-free case) is forbidden but shown here nevertheless as it corresponds to the origin of
the rotational band. In the simple model 1, terms off-diagonal in $v$ are neglected.
In model 2, the full matrix~\ref{Hnuc_mat2} is diagonalized. Values in parentheses are from Ref.~\cite{Turbiner:2018}.
}
\label{Table:rotv0}
\centering
\begin{tabular}{ccrrcccc}
\hline \hline
$L$  & Energy/$E_h$        &$M$ &$\pi$&  \multicolumn{4}{c}{Energy/$E_h$} \\ \hline
     & $B=0.0$ &    &     &  \multicolumn{2}{c}{$B=0.1$}&\multicolumn{2}{c}{$B=0.2$}\\
     &         &    &     & model 1 & model 2 & model 1 & model 2 \\ \hline
           &            &   -5 &    1 &  -1.15140 &  -1.15156 &  -1.13842 &  -1.13852 \\
           &            &    5 &    1 &  -1.15167 &  -1.15183 &  -1.13896 &  -1.13907 \\
           &            &   -4 &   -1 &  -1.15155 &  -1.15171 &  -1.13886 &  -1.13898 \\
           &            &    4 &   -1 &  -1.15177 &  -1.15193 &  -1.13929 &  -1.13941 \\
           &            &   -3 &    1 &  -1.15167 &  -1.15184 &  -1.13919 &  -1.13932 \\
$L=5$      & -1.15627   &    3 &    1 &  -1.15184 &  -1.15200 &  -1.13952 &  -1.13965 \\
           & (-1.15660) &   -2 &   -1 &  -1.15177 &  -1.15193 &  -1.13944 &  -1.13958 \\
           &            &    2 &   -1 &  -1.15188 &  -1.15204 &  -1.13966 &  -1.13979 \\
           &            &   -1 &    1 &  -1.15183 &  -1.15200 &  -1.13961 &  -1.13975 \\
           &            &    1 &    1 &  -1.15189 &  -1.15206 &  -1.13972 &  -1.13986 \\
           &            &    0 &   -1 &  -1.15187 &  -1.15204 &  -1.13970 &  -1.13984 \\
\hline
           &            &   -3 &    1 &  -1.15638 &  -1.15641 &  -1.14360 &  -1.14361 \\
           &            &    3 &    1 &  -1.15655 &  -1.15657 &  -1.14393 &  -1.14393 \\
           &            &   -2 &   -1 &  -1.15659 &  -1.15662 &  -1.14422 &  -1.14424 \\
$L=3$      & -1.16099   &    2 &   -1 &  -1.15670 &  -1.15672 &  -1.14444 &  -1.14445 \\
           & (-1.16130) &   -1 &    1 &  -1.15672 &  -1.15674 &  -1.14450 &  -1.14452 \\
           &            &    1 &    1 &  -1.15677 &  -1.15679 &  -1.14461 &  -1.14462 \\
           &            &    0 &   -1 &  -1.15677 &  -1.15680 &  -1.14466 &  -1.14468 \\
\hline
           &            &   -1 &    1 &  -1.15924 &  -1.15924 &  -1.14679 &  -1.14680 \\
$L=1$      & -1.16367   &    1 &    1 &  -1.15930 &  -1.15930 &  -1.14690 &  -1.14690 \\
           & (-1.16400) &    0 &   -1 &  -1.15959 &  -1.15959 &  -1.14783 &  -1.14783 \\
\hline
$L=0$      & -1.16421   &    0 &   -1 &  -1.15994 &  -1.15995 &  -1.14795 &  -1.14796 \\
           & (-1.16455) &      &      &          &          &           &          \\

\hline \hline
\end{tabular}
\end{table}

\begin{table}
\caption{Rotational energy levels of \hh in presence of a uniform magnetic field $B$ for the vibrational state $v=1$.
See Caption of Table~\ref{Table:rotv0} for explications.}
\label{Table:rotv1}
\centering
\begin{tabular}{ccrrcccc}
\hline \hline
$L$  & Energy/$E_h$        &$M$ &$\pi$&  \multicolumn{4}{c}{Energy/$E_h$} \\ \hline
     & $B=0.0$ &    &     &  \multicolumn{2}{c}{$B=0.1$}&\multicolumn{2}{c}{$B=0.2$}\\
     &         &    &     & model 1 & model 2 & model 1 & model 2 \\ \hline
           &            &   -4 &    1 &  -1.13514 &  -1.13521 &  -1.12176 &  -1.12179 \\
           &            &    4 &    1 &  -1.13536 &  -1.13543 &  -1.12220 &  -1.12222 \\
           &            &   -3 &   -1 &  -1.13534 &  -1.13541 &  -1.12235 &  -1.12239 \\
           &            &    3 &   -1 &  -1.13551 &  -1.13558 &  -1.12268 &  -1.12272 \\
$L=4$      & -1.14010   &   -2 &    1 &  -1.13549 &  -1.13556 &  -1.12272 &  -1.12277 \\
           & (-1.14050) &    2 &    1 &  -1.13560 &  -1.13567 &  -1.12294 &  -1.12299 \\
           &            &   -1 &   -1 &  -1.13558 &  -1.13566 &  -1.12297 &  -1.12302 \\
           &            &    1 &   -1 &  -1.13564 &  -1.13571 &  -1.12308 &  -1.12313 \\
           &            &    0 &    1 &  -1.13563 &  -1.13570 &  -1.12308 &  -1.12313 \\
\hline
           &            &   -2 &    1 &  -1.13888 &  -1.13888 &  -1.12575 &  -1.12575 \\
           &            &    2 &    1 &  -1.13898 &  -1.13899 &  -1.12597 &  -1.12597 \\
$L=2$      & -1.14364   &   -1 &   -1 &  -1.13918 &  -1.13918 &  -1.12666 &  -1.12666 \\
           & (-1.14405) &    1 &   -1 &  -1.13923 &  -1.13924 &  -1.12677 &  -1.12677 \\
           &            &    0 &    1 &  -1.13924 &  -1.13924 &  -1.12651 &  -1.12652 \\
\hline
$L=0$      & -1.14517   &    0 &    1 &  -1.14071 &  -1.14071 &  -1.12840 &  -1.12840 \\
           & (-1.14555)  &      &      &          &          &          &          \\

\hline \hline
\end{tabular}
\end{table}

\begin{table}
\caption{Rotational energy levels of \hh in presence of a uniform magnetic field $B$ for the vibrational state $v=2$. The pure vibrational state
  ($L=0$ in the field-free case) is forbidden but shown here nevertheless as it corresponds to the origin of the rotational band.
See Caption of Table \ref{Table:rotv0} for explications.}
\centering
\begin{tabular}{ccrrcccc}
\hline \hline
$L$  & Energy/$E_h$        &$M$ &$\pi$&  \multicolumn{4}{c}{Energy/$E_h$} \\ \hline
     & $B=0.0$ &    &     &  \multicolumn{2}{c}{$B=0.1$}&\multicolumn{2}{c}{$B=0.2$}\\
     &         &    &     & model 1 & model 2 & model 1 & model 2 \\ \hline
           &            &   -5 &    1 &  -1.11475 &  -1.11491 &  -1.10065 &  -1.10073 \\
           &            &    5 &    1 &  -1.11502 &  -1.11518 &  -1.10119 &  -1.10127 \\
           &            &   -4 &   -1 &  -1.11493 &  -1.11509 &  -1.10123 &  -1.10133 \\
           &            &    4 &   -1 &  -1.11515 &  -1.11531 &  -1.10167 &  -1.10176 \\
           &            &   -3 &    1 &  -1.11508 &  -1.11524 &  -1.10165 &  -1.10176 \\
$L=5$      & -1.12004   &    3 &    1 &  -1.11524 &  -1.11540 &  -1.10198 &  -1.10208 \\
           & (-1.12055) &   -2 &   -1 &  -1.11518 &  -1.11535 &  -1.10196 &  -1.10207 \\
           &            &    2 &   -1 &  -1.11529 &  -1.11546 &  -1.10217 &  -1.10229 \\
           &            &   -1 &    1 &  -1.11526 &  -1.11542 &  -1.10216 &  -1.10228 \\
           &            &    1 &    1 &  -1.11531 &  -1.11548 &  -1.10226 &  -1.10239 \\
           &            &    0 &   -1 &  -1.11530 &  -1.11547 &  -1.10226 &  -1.10238 \\
\hline
           &            &   -3 &    1 &  -1.11927 &  -1.11929 &  -1.10540 &  -1.10540 \\
           &            &    3 &    1 &  -1.11944 &  -1.11946 &  -1.10572 &  -1.10573 \\
           &            &   -2 &   -1 &  -1.11952 &  -1.11955 &  -1.10623 &  -1.10624 \\
$L=3$      & -1.12431   &    2 &   -1 &  -1.11963 &  -1.11966 &  -1.10645 &  -1.10646 \\
           & (-1.12475) &   -1 &    1 &  -1.11966 &  -1.11969 &  -1.10649 &  -1.10651 \\
           &            &    1 &    1 &  -1.11972 &  -1.11974 &  -1.10660 &  -1.10661 \\
           &            &    0 &   -1 &  -1.11973 &  -1.11976 &  -1.10672 &  -1.10674 \\
\hline
           &            &   -1 &    1 &  -1.12189 &  -1.12189 &  -1.10851 &  -1.10852 \\
$L=1$      & -1.12673   &    1 &    1 &  -1.12194 &  -1.12194 &  -1.10862 &  -1.10863 \\
           & (-1.12720) &    0 &   -1 &  -1.12232 &  -1.12232 &  -1.10986 &  -1.10986 \\
\hline
$L=0$      & -1.12722   &    0 &   -1 &  -1.12258 &  -1.12258 &  -1.10991 &  -1.10992 \\
           & (-1.12765) &      &      &          &          &           &          \\

\hline \hline
\end{tabular}
\label{Table:rotv2}
\end{table}

\begin{table}
\caption{Rotational energy levels of \hh in presence of a uniform magnetic field $B$ for the vibrational state $v=3$.
See Caption of Table \ref{Table:rotv0} for explications.}
\label{Table:rotv3}
\centering
\begin{tabular}{ccrrcccc}
\hline \hline
$L$  & Energy/$E_h$        &$M$ &$\pi$&  \multicolumn{4}{c}{Energy/$E_h$} \\ \hline
     & $B=0.0$ &    &     &  \multicolumn{2}{c}{$B=0.1$}&\multicolumn{2}{c}{$B=0.2$}\\
     &         &    &     & model 1 & model 2 & model 1 & model 2 \\ \hline
           &            &   -4 &    1 &  -1.10041 &  -1.10047 &  -1.08581 &  -1.08583 \\
           &            &    4 &    1 &  -1.10063 &  -1.10069 &  -1.08624 &  -1.08626 \\
           &            &   -3 &   -1 &  -1.10063 &  -1.10069 &  -1.08659 &  -1.08662 \\
           &            &    3 &   -1 &  -1.10079 &  -1.10085 &  -1.08692 &  -1.08694 \\
$L=4$      & -1.10550   &   -2 &    1 &  -1.10078 &  -1.10084 &  -1.08702 &  -1.08705 \\
           & (-1.10630) &    2 &    1 &  -1.10089 &  -1.10095 &  -1.08723 &  -1.08727 \\
           &            &   -1 &   -1 &  -1.10088 &  -1.10095 &  -1.08732 &  -1.08736 \\
           &            &    1 &   -1 &  -1.10094 &  -1.10100 &  -1.08742 &  -1.08747 \\
           &            &    0 &    1 &  -1.10093 &  -1.10100 &  -1.08743 &  -1.08748 \\
\hline
           &            &   -2 &    1 &  -1.10379 &  -1.10379 &  -1.08954 &  -1.08955 \\
           &            &    2 &    1 &  -1.10390 &  -1.10390 &  -1.08976 &  -1.08977 \\
$L=2$      & -1.10893   &   -1 &   -1 &  -1.10412 &  -1.10412 &  -1.09074 &  -1.09074 \\
           & (-1.10945) &    1 &   -1 &  -1.10417 &  -1.10417 &  -1.09085 &  -1.09085 \\
           &            &    0 &    1 &  -1.10417 &  -1.10418 &  -1.09031 &  -1.09033 \\
\hline
$L=0$      & -1.11034   &    0 &    1 &  -1.10550 &  -1.10550 &  -1.09251 &  -1.09251 \\
           & (-1.11085) &      &      &          &          &           &          \\
\hline \hline
\end{tabular}
\end{table}

\begin{table}
\caption{
Rotational energy levels of \dd in presence of a uniform magnetic field $B$ for the vibrational state $v=0$.
See Caption of Table \ref{Table:rotv0} for explications.}
\label{Table:Drotv0}
\centering
\begin{tabular}{ccrrcccc}
\hline \hline
$L$  & Energy/$E_h$        &$M$ &$\pi$&  \multicolumn{4}{c}{Energy/$E_h$} \\ \hline
     & $B=0.0$ &    &     &  \multicolumn{2}{c}{$B=0.1$}&\multicolumn{2}{c}{$B=0.2$}\\
     &         &    &     & model 1 & model 2 & model 1 & model 2 \\ \hline
           &            &   -4 &    1 &  -1.15985 &  -1.15987 &  -1.14715 &  -1.14715 \\
           &            &    4 &    1 &  -1.15996 &  -1.15998 &  -1.14737 &  -1.14737 \\
           &            &   -3 &   -1 &  -1.16001 &  -1.16003 &  -1.14764 &  -1.14765 \\
           &            &    3 &   -1 &  -1.16009 &  -1.16011 &  -1.14780 &  -1.14781 \\
$L=4$      & -1.158877  &   -2 &    1 &  -1.16012 &  -1.16013 &  -1.14790 &  -1.14791 \\
           &            &    2 &    1 &  -1.16017 &  -1.16019 &  -1.14800 &  -1.14802 \\
           &            &   -1 &   -1 &  -1.16019 &  -1.16020 &  -1.14808 &  -1.14809 \\
           &            &    1 &   -1 &  -1.16021 &  -1.16023 &  -1.14813 &  -1.14815 \\
           &            &    0 &    1 &  -1.16022 &  -1.16024 &  -1.14814 &  -1.14816 \\
\hline
           &            &   -2 &    1 &  -1.16185 &  -1.16185 &  -1.14936 &  -1.14936 \\
           &            &    2 &    1 &  -1.16190 &  -1.16190 &  -1.14947 &  -1.14947 \\
$L=2$      & -1.162594  &   -1 &   -1 &  -1.16209 &  -1.16209 &  -1.15012 &  -1.15012 \\
           &            &    1 &   -1 &  -1.16212 &  -1.16212 &  -1.15017 &  -1.15017 \\
           &            &    0 &    1 &  -1.16211 &  -1.16211 &  -1.14981 &  -1.14982 \\
\hline
$L=0$      & -1.164212  &    0 &    1 &  -1.16291 &  -1.16291 &  -1.15118 &  -1.15119 \\
\hline \hline
\end{tabular}
\end{table}

\begin{table}
\caption{Rotational energy levels of \dd in presence of a uniform magnetic field $B$ for the vibrational state $v=1$. The pure vibrational state
  ($L=0$ in the field-free case) is forbidden but shown here nevertheless as it corresponds to the origin of the rotational band.
See Caption of Table \ref{Table:rotv0} for explications.}
\label{Table:Drotv1}
\centering
\begin{tabular}{ccrrcccc}
\hline \hline
$L$  & Energy/$E_h$        &$M$ &$\pi$&  \multicolumn{4}{c}{Energy/$E_h$} \\ \hline
     & $B=0.0$ &    &     &  \multicolumn{2}{c}{$B=0.1$}&\multicolumn{2}{c}{$B=0.2$}\\
     &         &    &     & model 1 & model 2 & model 1 & model 2 \\ \hline
           &            &   -5 &    1 &  -1.14472 &  -1.14475 &  -1.13162 &  -1.13163 \\
           &            &    5 &    1 &  -1.14485 &  -1.14489 &  -1.13189 &  -1.13190 \\
           &            &   -4 &   -1 &  -1.14487 &  -1.14491 &  -1.13208 &  -1.13210 \\
           &            &    4 &   -1 &  -1.14498 &  -1.14501 &  -1.13230 &  -1.13232 \\
           &            &   -3 &    1 &  -1.14499 &  -1.14502 &  -1.13238 &  -1.13241 \\
$L=5$      & -1.137626  &    3 &    1 &  -1.14507 &  -1.14511 &  -1.13255 &  -1.13257 \\
           &            &   -2 &   -1 &  -1.14507 &  -1.14511 &  -1.13260 &  -1.13263 \\
           &            &    2 &   -1 &  -1.14513 &  -1.14516 &  -1.13271 &  -1.13274 \\
           &            &   -1 &    1 &  -1.14513 &  -1.14517 &  -1.13274 &  -1.13277 \\
           &            &    1 &    1 &  -1.14515 &  -1.14519 &  -1.13279 &  -1.13282 \\
           &            &    0 &   -1 &  -1.14515 &  -1.14519 &  -1.13280 &  -1.13283 \\
\hline
           &            &   -3 &    1 &  -1.14716 &  -1.14716 &  -1.13423 &  -1.13423 \\
           &            &    3 &    1 &  -1.14724 &  -1.14725 &  -1.13439 &  -1.13439 \\
           &            &   -2 &   -1 &  -1.14738 &  -1.14738 &  -1.13490 &  -1.13490 \\
$L=3$      & -1.142112  &    2 &   -1 &  -1.14743 &  -1.14744 &  -1.13501 &  -1.13501 \\
           &            &   -1 &    1 &  -1.14748 &  -1.14749 &  -1.13496 &  -1.13497 \\
           &            &    1 &    1 &  -1.14751 &  -1.14752 &  -1.13502 &  -1.13502 \\
           &            &    0 &   -1 &  -1.14753 &  -1.14754 &  -1.13521 &  -1.13522 \\
\hline
           &            &   -1 &    1 &  -1.14862 &  -1.14862 &  -1.13616 &  -1.13616 \\
$L=1$      & -1.144658  &    1 &    1 &  -1.14864 &  -1.14864 &  -1.13621 &  -1.13622 \\
           &            &    0 &   -1 &  -1.14899 &  -1.14899 &  -1.13715 &  -1.13715 \\
\hline
$L=0$      & -1.145172  &    0 &   -1 &  -1.14908 &  -1.14908 &  -1.13716 &  -1.13716 \\
\hline \hline
\end{tabular}
\end{table}

\begin{table}
\caption{Rotational energy levels of \dd in presence of a uniform magnetic field $B$ for the vibrational state $v=2$.
See Caption of Table \ref{Table:rotv0} for explications.}
\label{Table:Drotv2}
\centering
\begin{tabular}{ccrrcccc}
\hline \hline
$L$  & Energy/$E_h$        &$M$ &$\pi$&  \multicolumn{4}{c}{Energy/$E_h$} \\ \hline
     & $B=0.0$ &    &     &  \multicolumn{2}{c}{$B=0.1$}&\multicolumn{2}{c}{$B=0.2$}\\
     &         &    &     & model 1 & model 2 & model 1 & model 2 \\ \hline
           &            &   -4 &    1 &  -1.13286 &  -1.13288 &  -1.11944 &  -1.11944 \\
           &            &    4 &    1 &  -1.13297 &  -1.13299 &  -1.11965 &  -1.11966 \\
           &            &   -3 &   -1 &  -1.13306 &  -1.13307 &  -1.12005 &  -1.12006 \\
           &            &    3 &   -1 &  -1.13314 &  -1.13315 &  -1.12022 &  -1.12022 \\
$L=4$      & -1.122394  &   -2 &    1 &  -1.13318 &  -1.13320 &  -1.12030 &  -1.12032 \\
           &            &    2 &    1 &  -1.13324 &  -1.13325 &  -1.12041 &  -1.12042 \\
           &            &   -1 &   -1 &  -1.13326 &  -1.13328 &  -1.12054 &  -1.12055 \\
           &            &    1 &   -1 &  -1.13329 &  -1.13331 &  -1.12059 &  -1.12060 \\
           &            &    0 &    1 &  -1.13330 &  -1.13332 &  -1.12058 &  -1.12060 \\
\hline
           &            &   -2 &    1 &  -1.13474 &  -1.13474 &  -1.12161 &  -1.12162 \\
           &            &    2 &    1 &  -1.13480 &  -1.13480 &  -1.12172 &  -1.12173 \\
$L=2$      & -1.125757  &   -1 &   -1 &  -1.13504 &  -1.13504 &  -1.12254 &  -1.12254 \\
           &            &    1 &   -1 &  -1.13507 &  -1.13507 &  -1.12259 &  -1.12259 \\
           &            &    0 &    1 &  -1.13503 &  -1.13503 &  -1.12204 &  -1.12206 \\
\hline
$L=0$      & -1.127217  &    0 &    1 &  -1.13582 &  -1.13582 &  -1.12367 &  -1.12367 \\
\hline \hline
\end{tabular}
\end{table}

\begin{table}
\caption{Rotational energy levels of \dd in presence of a uniform magnetic field $B$ for the vibrational state $v=3$. The pure vibrational state
  ($L=0$ in the field-free case) is forbidden but shown here nevertheless as it corresponds to the origin of the rotational band.
See Caption of Table \ref{Table:rotv0} for explications.}
\label{Table:Drotv3}
\centering
\begin{tabular}{ccrrcccc}
\hline \hline
$L$  & Energy/$E_h$        &$M$ &$\pi$&  \multicolumn{4}{c}{Energy/$E_h$} \\ \hline
     & $B=0.0$ &    &     &  \multicolumn{2}{c}{$B=0.1$}&\multicolumn{2}{c}{$B=0.2$}\\
     &         &    &     & model 1 & model 2 & model 1 & model 2 \\ \hline
           &            &   -5 &    1 &  -1.11895 &  -1.11898 &  -1.10501 &  -1.10502 \\
           &            &    5 &    1 &  -1.11908 &  -1.11912 &  -1.10529 &  -1.10529 \\
           &            &   -4 &   -1 &  -1.11912 &  -1.11915 &  -1.10559 &  -1.10560 \\
           &            &    4 &   -1 &  -1.11923 &  -1.11926 &  -1.10581 &  -1.10582 \\
           &            &   -3 &    1 &  -1.11924 &  -1.11928 &  -1.10593 &  -1.10595 \\
$L=5$      & -1.102915  &    3 &    1 &  -1.11933 &  -1.11936 &  -1.10609 &  -1.10611 \\
           &            &   -2 &   -1 &  -1.11934 &  -1.11938 &  -1.10619 &  -1.10621 \\
           &            &    2 &   -1 &  -1.11939 &  -1.11943 &  -1.10630 &  -1.10632 \\
           &            &   -1 &    1 &  -1.11940 &  -1.11944 &  -1.10634 &  -1.10636 \\
           &            &    1 &    1 &  -1.11942 &  -1.11946 &  -1.10639 &  -1.10642 \\
           &            &    0 &   -1 &  -1.11943 &  -1.11947 &  -1.10641 &  -1.10644 \\
\hline
           &            &   -3 &    1 &  -1.12123 &  -1.12124 &  -1.10752 &  -1.10752 \\
           &            &    3 &    1 &  -1.12131 &  -1.12132 &  -1.10768 &  -1.10769 \\
           &            &   -2 &   -1 &  -1.12148 &  -1.12148 &  -1.10836 &  -1.10836 \\
$L=3$      & -1.107489  &    2 &   -1 &  -1.12153 &  -1.12154 &  -1.10847 &  -1.10847 \\
           &            &   -1 &    1 &  -1.12158 &  -1.12159 &  -1.10828 &  -1.10829 \\
           &            &    1 &    1 &  -1.12161 &  -1.12162 &  -1.10833 &  -1.10835 \\
           &            &    0 &   -1 &  -1.12164 &  -1.12164 &  -1.10867 &  -1.10868 \\
\hline
           &            &   -1 &    1 &  -1.12262 &  -1.12262 &  -1.10960 &  -1.10960 \\
$L=1$      & -1.109872  &    1 &    1 &  -1.12264 &  -1.12264 &  -1.10966 &  -1.10966 \\
           &            &    0 &   -1 &  -1.12303 &  -1.12303 &  -1.11073 &  -1.11073 \\
\hline
$L=0$      & -1.110336  &    0 &   -1 &  -1.12310 &  -1.12310 &  -1.10896 &  -1.10898 \\
\hline \hline
\end{tabular}
\end{table}

\clearpage

\begin{figure}[htp!]
\centering
\includegraphics[angle=-90, width=0.8\textwidth]{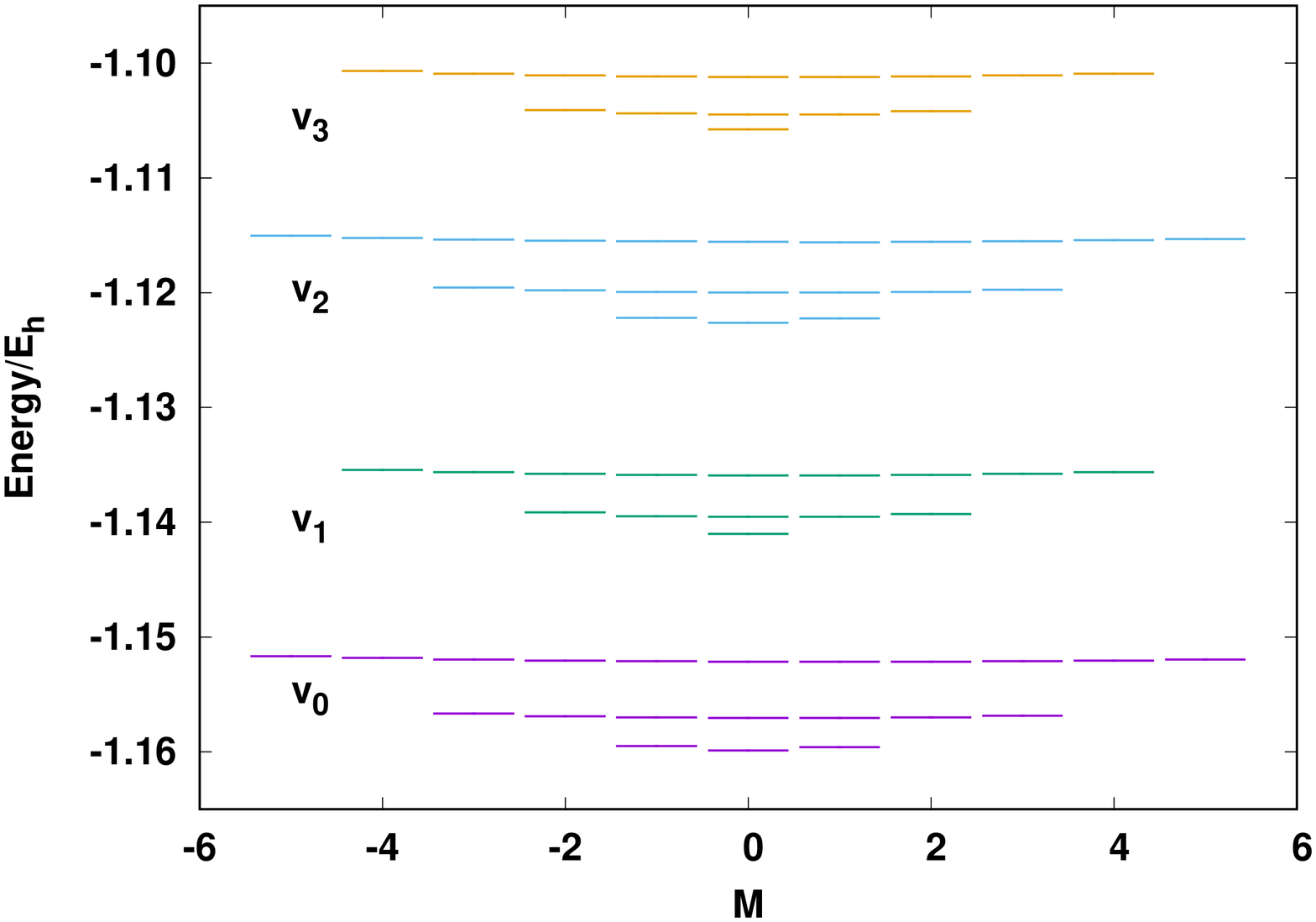}
\caption{Rotational structure, up to $L=5$, of the four lowest vibrational
states of \hh in the presence of an external magnetic field $B=0.1\, B_0$.
}
\label{rovibplotB01}
\end{figure}

\begin{figure}[htp!]
\centering
\includegraphics[angle=-90, width=0.8\textwidth]{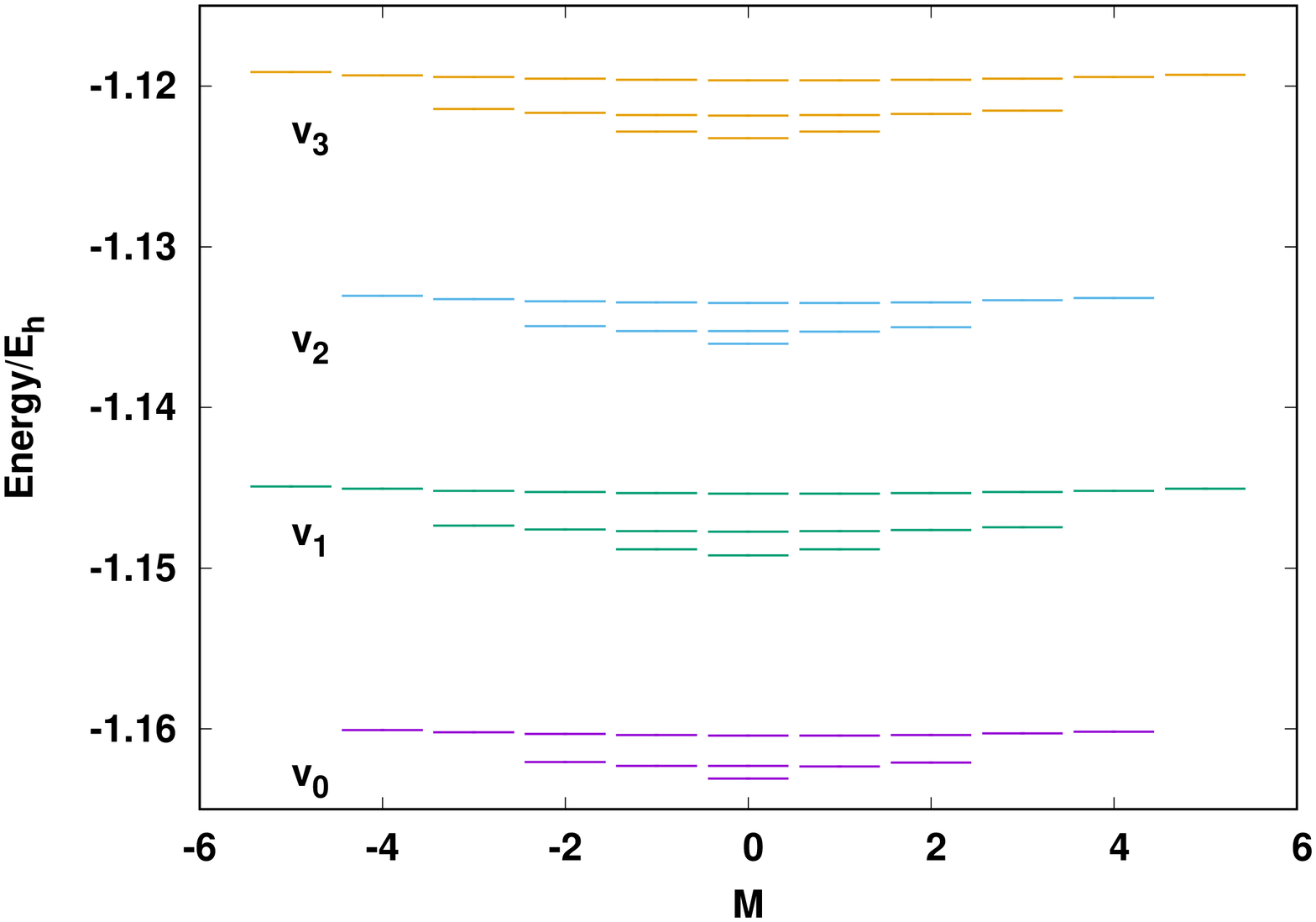}
\caption{Rotational structure, up to $L=5$, of the four lowest vibrational
states of \dd in the presence of an external magnetic field $B=0.1\, B_0$.
}
\label{DrovibplotB01}
\end{figure}

\section{Conclusions}
We have investigated the problem of the hydrogen molecule vibrating and
rotating in the presence of an external magnetic field for the field strengths
of $B = 0.1, 0.15, 0.175~\rm a.u.$ and $B = 0.2~\rm
a.u.$ ($4.7\times 10^4~ \rm T$). It was shown that for $B > B_{cr} =
0.178$\,a.u. H$_2$ exists in the form of two isolated hydrogen atoms with
anti-parallel electron spins to the magnetic field direction. For magnetic
fields larger than $12$\,a.u. the molecule gets bound in parallel
configuration with ${}^3\Pi_u$ as the ground state, see e.g. 
\cite{Detmer:1997,Detmer:1998a}.

Highly accurate variational calculations, based on a few-parameter physically
adequate trial function, are carried out for
inclined configurations, where the molecular axis forms an angle $\theta$ with
respect to the direction of a uniform constant magnetic field.
We calculated diamagnetic and paramagnetic susceptibilities (for $\theta=45^\circ$ for the
first time), they closely describe experimental data and agree very well
with other calculations or are superior.
The two-dimensional potential energy surfaces were built for magnetic fields 
for $B=0.1$ and $0.2$\,a.u. The parallel orientation of the H$_2$ molecule 
with respect to the magnetic field is the most stable one even though the molecule 
becomes metastable for $B=0.2$\,a.u. being in domain $B > B_{cr}$. This holds true 
also if the vibrational zero-point energy is taken into account. 
Though the rovibrational ground state is located well
above the barrier to perpendicular orientation, the vibrating molecule remains
in its parallel orientation. The lowest rovibrational states have then been
calculated for the first time. Their energy values are reported for the four
lowest vibrational states and rotational excitation up to $M = 5$, for both the
H$_2$ and D$_2$ isotopologues.

\section{Acknowledgements}
The authors thank the high-performance computer centre ROMEO of the University
of Reims Champagne-Ardenne, CRIANN of the Region of Normandy, France and
cluster KAREN (ICN-UNAM, Mexico) for generous allowance of super-computer time.
The research by J.C.L.V., D.J.N., A.V.T. is partially supported by CONACyT grant 
A1-S-17364 and DGAPA grant IN108815 (Mexico).
This work was also supported by the Programme National de Plan\'etologie
(PNP) of CNRS/INSU, co-funded by CNES.
Two of the authors A.A. and A.V.T. have the honor and the privilege to know
closely Vladimir Tyuterev to whom this paper is dedicated.




\begin{thebibliography}{00}
\expandafter\ifx\csname url\endcsname\relax
  \def\url#1{\texttt{#1}}\fi
\expandafter\ifx\csname urlprefix\endcsname\relax\def\urlprefix{URL }\fi
\expandafter\ifx\csname href\endcsname\relax
  \def\href#1#2{#2} \def\path#1{#1}\fi

\bibitem{Woltjer:1964}
L.~{Woltjer}, X-rays and type {I} supernova remnants., Astrophys. J. 140 (1964)
  1309--1313.
\newblock \href {http://dx.doi.org/10.1086/148028} {\path{doi:10.1086/148028}}.

\bibitem{Pacini:1967}
F.~{Pacini}, Energy emission from a neutron star, Nature 216 (1967) 567--568.
\newblock \href {http://dx.doi.org/10.1038/216567a0}
  {\path{doi:10.1038/216567a0}}.

\bibitem{Gold:1968}
T.~{Gold}, Rotating neutron stars as the origin of the pulsating radio sources,
  Nature 218 (1968) 731--732.
\newblock \href {http://dx.doi.org/10.1038/218731a0}
  {\path{doi:10.1038/218731a0}}.

\bibitem{Goldreich:1969}
P.~{Goldreich}, W.~H. {Julian}, Pulsar electrodynamics, Astrophys. J. 157
  (1969) 869.
\newblock \href {http://dx.doi.org/10.1086/150119} {\path{doi:10.1086/150119}}.

\bibitem{Garcia:2016}
E.~Garc\'ia-Berro, M.~Kilic, S.~O. Kepler, Magnetic white dwarfs: Observations,
  theory and future prospects, Int. J. Mod. Phys. D 25 (2016) 1630005.
\newblock \href {http://dx.doi.org/10.1142/S0218271816300056}
  {\path{doi:10.1142/S0218271816300056}}.

\bibitem{Ruderman:1971}
M.~A. Ruderman, Matter in superstrong magnetic fields: the surface of a neutron
  star, Phys. Rev. Lett. 27 (1971) 1306--1308.
\newblock \href {http://dx.doi.org/10.1103/PhysRevLett.27.1306}
  {\path{doi:10.1103/PhysRevLett.27.1306}}.

\bibitem{Kadomtsev:1971a}
B.~B. Kadomtsev, V.~S. Kudryavtsev, Molecules in an ultrastrong magnetic field,
  {Pis'ma} Zh. Eksp. Teor. Fiz. [Sov. Phys. - JETP Lett.] 13 (1971) 15--19, {
  Sov. Phys. - JETP Lett. \bf 13} (1971) 9-12 (English Translation).

\bibitem{Kadomtsev:1972}
B.~B. Kadomtsev, V.~S. Kudryavtsev, Matter in a superstrong magnetic field, Zh.
  Eksp. Teor. Fiz. [Sov. Phys. - JETP] 62 (1972) 144, {Sov. Phys. - JETP \bf
  35} (1972) 76-80 (English Translation).

\bibitem{Kravchenko:1996}
Y.~P. Kravchenko, M.~A. Liberman, B.~Johansson, Exact solution for a hydrogen
  atom in a magnetic field of arbitrary strength, Phys. Rev. A 54 (1996)
  287--305.
\newblock \href {http://dx.doi.org/10.1103/PhysRevA.54.287}
  {\path{doi:10.1103/PhysRevA.54.287}}.

\bibitem{Turbiner:2006PR}
A.~V. Turbiner, J.~C. {L\'opez Vieyra}, One-electron molecular systems in a
  strong magnetic field, Physics Reports 424 (2006) 309--396.
\newblock \href {http://dx.doi.org/10.1016/j.physrep.2005.11.002}
  {\path{doi:10.1016/j.physrep.2005.11.002}}.

\bibitem{Turbiner:1983}
A.~V. Turbiner, Hydrogen molecule in a strong magnetic field, {Pis'ma} Zh.
  Eksp. Teor. Fiz. [Sov. Phys. - JETP Lett.] 38 (1983) 510--514, {JETP Lett.
  \bf 38} (1983) 618-622 (English Translation).

\bibitem{Turbiner:2007PRA}
A.~V. {Turbiner}, N.~L. {Guevara}, J.~C. {L{\'o}pez Vieyra}, {H$_3^+$ molecular
  ion in a magnetic field: Linear parallel configuration}, Phys. Rev. A 75
  (2007) 053408.
\newblock \href {http://arxiv.org/abs/physics/0606083}
  {\path{arXiv:physics/0606083}}, \href
  {http://dx.doi.org/10.1103/PhysRevA.75.053408}
  {\path{doi:10.1103/PhysRevA.75.053408}}.

\bibitem{MED13:9871}
H.~Medel~Cobaxin, A.~Alijah,
  \href{http://pubs.acs.org/doi/abs/10.1021/jp312856s}{{Vibrating H$_3^+$} in a
  uniform magnetic field}, J. Phys. Chem. A 117~(39) (2013) 9871--9881.
\newblock \href
  {http://arxiv.org/abs/http://pubs.acs.org/doi/pdf/10.1021/jp312856s}
  {\path{arXiv:http://pubs.acs.org/doi/pdf/10.1021/jp312856s}}, \href
  {http://dx.doi.org/10.1021/jp312856s} {\path{doi:10.1021/jp312856s}}.
\newline\urlprefix\url{http://pubs.acs.org/doi/abs/10.1021/jp312856s}

\bibitem{Turbiner:2010}
A.~V. Turbiner, J.~C. L\'opez~Vieyra, N.~L. Guevara, Charged hydrogenic,
  helium, and helium-hydrogenic molecular chains in a strong magnetic field,
  Phys. Rev. A 81 (2010) 042503.
\newblock \href {http://dx.doi.org/10.1103/PhysRevA.81.042503}
  {\path{doi:10.1103/PhysRevA.81.042503}}.

\bibitem{Medel:2015}
H.~{Medel Cobaxin}, A.~{Alijah}, J.~C. {L{\'o}pez Vieyra}, A.~V. {Turbiner},
  {H$_{2}$$^{+}$ in a weak magnetic field}, J. Phys. B 48.
\newblock \href {http://dx.doi.org/10.1088/0953-4075/48/4/045101}
  {\path{doi:10.1088/0953-4075/48/4/045101}}.

\bibitem{Detmer:1997}
T.~Detmer, P.~Schmelcher, F.~K. Diakonos, L.~S. Cederbaum, Hydrogen molecule in
  magnetic fields: The ground states of the \ensuremath{\Sigma}-manifold of the
  parallel configuration, Phys. Rev. A 56 (1997) 1825--1838.
\newblock \href {http://dx.doi.org/10.1103/PhysRevA.56.1825}
  {\path{doi:10.1103/PhysRevA.56.1825}}.

\bibitem{Detmer:1998a}
T.~Detmer, P.~Schmelcher, L.~S. Cederbaum, Hydrogen molecule in a magnetic
  field: The lowest states of the $\ensuremath{\Pi}$-manifold and the global
  ground state of the parallel configuration, Phys. Rev. A 57 (1998)
  1767--1777.
\newblock \href {http://dx.doi.org/10.1103/PhysRevA.57.1767}
  {\path{doi:10.1103/PhysRevA.57.1767}}.

\bibitem{Turbiner:1984}
A.~V. Turbiner, The eigenvalue spectrum in quantum mechanics and the
  nonlinearization procedure, Usp. Fiz. Nauk. 144 (1984) 35--78, {\it Soviet
  Phys. -- Uspekhi \bf 27} (1984) 668 (English Translation).

\bibitem{Turbiner:2007}
A.~V. Turbiner, N.~L. Guevara, A note about the ground state of the hydrogen
  molecule, Collect. Czech. Chem. Commun. 72 (2007) 164--170.
\newblock \href {http://dx.doi.org/https://doi.org/10.1135/cccc20070164}
  {\path{doi:https://doi.org/10.1135/cccc20070164}}.

\bibitem{GenzMalik1980}
A.~Genz, A.~Malik, Remarks on algorithm 006: An adaptive algorithm for
  numerical integration over an {$N$}-dimensional rectangular region, J.
  Comput. Appl. Math. 6 (1980) 295--302.
\newblock \href
  {http://dx.doi.org/https://doi.org/10.1016/0771-050X(80)90039-X}
  {\path{doi:https://doi.org/10.1016/0771-050X(80)90039-X}}.

\bibitem{Sims:2006}
J.~S. Sims, S.~A. Hagstrom, High precision variational calculations for the
  {B}orn-{O}ppenheimer energies of the ground state of the hydrogen molecule,
  J. Chem. Phys. 124 (2006) 094101.
\newblock \href {http://dx.doi.org/10.1063/1.2173250}
  {\path{doi:10.1063/1.2173250}}.

\bibitem{Turbiner:2018}
H.~{Olivares-Pil{\'o}n}, A.~V. {Turbiner}, {H$_{2}$$^{+}$, HeH and H$_{2}$:
  Approximating potential curves, calculating rovibrational states}, Annals of
  Physics 393 (2018) 335--357.
\newblock \href {http://dx.doi.org/10.1016/j.aop.2018.04.021}
  {\path{doi:10.1016/j.aop.2018.04.021}}.

\bibitem{Potekhin:2001}
A.~Y. Potekhin, A.~V. Turbiner, Hydrogen atom in a magnetic field: The
  quadrupole moment, Phys. Rev. A 63 (2001) 065402.
\newblock \href {http://dx.doi.org/10.1103/PhysRevA.63.065402}
  {\path{doi:10.1103/PhysRevA.63.065402}}.

\bibitem{Ramsey}
N.~Ramsey, \href{https://books.google.fr/books?id=T\_7Hg08X7CMC}{Molecular
  Beams}, International series of monographs on physics, Oxford University
  Press, Oxford, 1956.
\newline\urlprefix\url{https://books.google.fr/books?id=T\_7Hg08X7CMC}

\bibitem{Riley:1977}
J.~P. Riley, W.~T. Raynes, The octopole magnetizability of the hydrogen
  molecule, Molecular Physics 33 (1977) 631--634.
\newblock \href {http://dx.doi.org/10.1080/00268977700100581}
  {\path{doi:10.1080/00268977700100581}}.

\bibitem{KOLOS:1965}
W.~Kolos, L.~Wolniewicz, Potential-energy curves for the {$X {}^1\Sigma_g^+$,
  $b{}^3\Sigma_u^+$, and $C{}^1\Pi_u$} states of the hydrogen molecule, J.
  Chem. Phys. 43 (1965) 2429--2441.
\newblock \href {http://dx.doi.org/10.1063/1.1697142}
  {\path{doi:10.1063/1.1697142}}.

\bibitem{MARUANI}
J.~Rychlewski, Electric and magnetic properties for the ground and excited
  states of molecular hydrogen, in: J.~Maruani (Ed.), {M}olecules in Physics,
  Chemistry, and Biology. Vol. II. Physical aspects of molecular systems,
  Kluwer Academic Publishers, (1988), pp. 207--255.

\bibitem{Pflug:1977}
D.~R. {Pflug}, W.~E. {Palke}, B.~{Kirtman}, {Calculation of the magnetic
  susceptibility of H$_{2}$ by the distinguishable electron method}, J. Chem.
  Phys. 67 (1977) 1676--1683.
\newblock \href {http://dx.doi.org/10.1063/1.435000}
  {\path{doi:10.1063/1.435000}}.

\end{thebibliography}






\end{document}